\documentclass[iop]{emulateapj}
\usepackage{hyperref}
\usepackage{amsmath}
\usepackage[flushleft]{threeparttable}

\newcommand{\tmb}{$T_{\rm mb}~$}

\newcommand{\tex}{$T_{\rm ex}~$}

\newcommand{\ddfrac}{$D_{\rm frac}^{\rm N_2H^+}~$}

\newcommand{\ntdp}{$\rm N_2D^+~$}
\newcommand{\ntdpns}{$\rm N_2D^+$}

\newcommand{\dcopns}{$\rm DCO^+$}
\newcommand{\ceio}{$\rm C^{18}O~$}
\newcommand{\ceions}{$\rm C^{18}O$}

\newcommand{\nthp}{$\rm N_2H^+~$}
\newcommand{\nthpns}{$\rm N_2H^+$}
\newcommand{\kms}{$\rm km~s^{-1}~$}
\newcommand{\kmsns}{$\rm km~s^{-1}$}
\newcommand{\jyb}{$\rm Jy\:beam^{-1}\:$}
\newcommand{\jybns}{$\rm Jy\:beam^{-1}$}

\shorttitle{Zooming in to Massive Star Birth}
\shortauthors{Kong et al.}

\begin{document}

\title{Zooming in to Massive Star Birth}

\author{Shuo Kong\altaffilmark{1}}
\affil{Dept. of Astronomy, Yale University, New Haven, Connecticut 06511, USA}

\author{Jonathan C. Tan\altaffilmark{2,3}}
\affil{Dept. of Space, Earth and Environment, Chalmers University of Technology, Gothenburg, Sweden}
\affil{Dept. of Astronomy, University of Virginia, Charlottesville, Virginia 22904, USA}

\author{Paola Caselli\altaffilmark{4}}
\affil{Max-Planck-Institute for Extraterrestrial Physics (MPE), Giessenbachstr. 1, D-85748 Garching, Germany}

\author{Francesco Fontani\altaffilmark{5}}
\affil{INAF - Osservatorio Astrofisico di Arcetri, L.go E. Fermi 5 I-50125, Florence, Italy}

\author{Ke Wang\altaffilmark{6}}
\affil{European Southern Observatory (ESO), Karl-Schwarzschild-Str. 2, D-85748 Garching, Germany}

\author{Michael J. Butler\altaffilmark{7}}
\affil{Max Planck Institute for Astronomy, K\"onigstuhl 17, D-69117 Heidelberg, Germany}

\begin{abstract}
We present high resolution (0.2\arcsec, 1000~AU) 1.3~mm {\it ALMA}
observations of massive infrared dark cloud clump, G028.37+00.07-C1,
thought to harbor the early stages of massive star formation. Using
\ntdpns(3-2) we resolve the previously identified C1-S core,
separating the bulk of its emission from two nearby protostellar
sources. C1-S is thus identified as a massive ($\sim50\:M_\odot$),
compact ($\sim0.1\:$pc diameter) starless core, e.g., with no
signs of outflow activity. Being highly deuterated, this is a
promising candidate for a pre-stellar core on the verge of
collapse. An analysis of its dynamical state indicates a sub-virial
velocity dispersion compared to a trans-Alfv\'enic turbulent core
model. However, virial equilibrium could be achieved with
sub-Alfv\'enic conditions involving $\sim2\:$mG magnetic field
strengths.
\end{abstract}


\section{Introduction}

One way to distinguish between different theoretical models of massive
star formation is to study the initial conditions of gas before
formation of a massive protostar \citep[see, e.g.,][for a
  review]{2014prpl.conf..149T}. Core accretion models, e.g., the
Turbulent Core model of \citet[][hereafter
  MT03]{2002Natur.416...59M,2003ApJ...585..850M}, assume that this
initial condition is a massive pre-stellar core (PSC).  By analogy
with low-mass PSCs \citep[see, e.g.,][]{2012A&ARv..20...56C,2014ApJ...797...27F}, 
these cores are expected to be highly deuterated and thus well-traced
by species such as \ntdpns. On the other hand, competitive
  accretion models \citep[e.g.,][]{2001MNRAS.323..785B,2010ApJ...709...27W} do
  not involve massive PSCs as the initial conditions for massive star
  formation. 

\citet[][hereafter T13]{2013ApJ...779...96T} used \ntdpns(3-2)
to identify the C1-S core in a pilot survey with ALMA in Cycle 0
at 2.3\arcsec\ resolution of four infrared dark cloud (IRDC) clumps
\citep[see also][for an extension of this survey to 32 more
IRDC clumps]{2017ApJ...834..193K}.
The C1-S core is at the western end of the massive IRDC
G028.37+00.07 at a distance of about 5~kpc \citep{2006ApJ...641..389R}. 
\citet{2016ApJ...821...94K} measured the deuteration fraction,
$D_{\rm frac}^{\rm N2H+}\equiv [{\rm N_2D^+}]/[{\rm N_2H^+}]$, in this
``core'' (i.e., roughly 3.5\arcsec in radius) to be $\simeq0.2$ to
0.7, which, by comparison with the models of
\citet{2015ApJ...804...98K}, may indicate a relatively old
astrochemical age compared to the local free-fall time of the
core. 

\citet[][hereafter T16]{2016ApJ...821L...3T} presented results from
the compact configuration (i.e., 1.5\arcsec resolution) follow-up of
the C1-S region with ALMA in Cycle 2, in particular reporting the
detection of two protostellar outflow sources in the vicinity (i.e.,
within $\sim0.1\:$pc) of the core center \citep[see
  also][]{2016ApJ...828..100F}.  If these protostars are embedded
  in the core, then this would obviously invalidate its status as a
  PSC.  Such a situation arose for the massive IR-quiet core N63
\citep{2010A&A...524A..18B}, which was later observed to host a powerful,
collimated outflow \citep{2013A&A...558A.125D}.
 The only other previously reported candidate massive PSC is
  G11.92−0.61-MM2, detected solely by its continuum emission
  \citep{2014ApJ...796L...2C}. However, the non-detection of any
  molecular lines toward this source is peculiar and makes it
  difficult to assess the reliability of the structure, e.g., via a
  dynamical mass measurement.

Here we present the full results of our Cycle 2 observations of C1-S,
combining compact and extended configuration observations that achieve
0.2\arcsec resolution.  With this higher-resolution data, we are
  able to spatially and kinematically resolve the \ntdpns(3-2) core
  from the nearby protostars, so we conclude it remains a promising
  candidate to be a massive PSC.

\section{Observations}\label{sec:obs}

\subsection{ALMA Observations}\label{sec:obsalma}

The observations were carried out during {\it ALMA} Cycle 2, under the
project 2013.1.00248.S (PI: Tan) using two configurations of the
12m-array. The compact configuration (C34-1) 
observation was performed on
05-Apr-2015 (UTC) with 34 12-m antennas and a total on-source  integration time
of 2087  seconds.  
 The baselines range from 15m to 327m, corresponding to angular
scales from 10\arcsec~to 1\arcsec.
The angular resolution obtained from this
observation is about 1.5\arcsec .  
The bandpass calibrator was J1924-2914;
the flux calibrator was Neptune; the gain calibrator was J1912-0804;
and the delay calibrator was J1902-0458. 
 The system temperature was about 85 K.
The extended configuration (C34-6) 
data were taken on 03-Jul-2015 (UTC) with 43 12-m antennas and a total
 on-source integration time of 4174  seconds.  
 The baselines range from 34m to 1574m, corresponding to angular
scales from 2.3\arcsec~to 0.2\arcsec.
The angular resolution obtained
from this observation is about 0.25\arcsec .  
The bandpass calibrator was
J1751+0939; the flux calibrator was Titan; the gain calibrator was
J1827-0405; and the delay calibrator was J1912-0804.  
 The system temperature was about 95 K.
Both observations targeted R.A.=$\rm 18^h42^m46\fs585634$, DEC.=$\rm
-04\arcdeg04\arcmin12\farcs36111$ (J2000), which is inbetween the C1-N
and C1-S (see T16). The primary beam FWHM was $\sim$ 26\arcsec, 
covering both sources.

The observations were carried out in {\it ALMA} band 6.  Four
basebands and seven spectral windows were used. Baseband 1 was set to
a single spectral window centered on the source frame \ntdpns(3-2)
line (rest frequency is 231.32183~GHz; C1 source frame radial velocity is
79.4$\:{\rm km\:s}^{-1}$), with a total bandwidth of 58 MHz (76 \kmsns) 
and a velocity resolution of 0.05 \kmsns.
Baseband 2 was set to a single spectral window for a continuum
observation, centered at 231.00~GHz, with a total bandwidth of about
2~GHz, but also including coverage of $^{12}$CO(2-1) (rest frequency
230.538~GHz) with a velocity resolution of 1.3~\kmsns. This set-up is
especially useful for identifying CO outflows from protostars. 
 (The CO line is excluded from the continuum band and
the final aggregate continuum bandwidth is $\sim$ 1.4 GHz for imaging). 
Baseband 3 was set to a single spectral window centered on
\ceions(2-1) (rest frequency 219.56036~GHz) with 
 a total bandwidth of 58 MHz (80 \kmsns) 
and a velocity resolution
of 0.05 \kmsns. Baseband 4 was split into four spectral windows at
rest frequencies of: 216.11258~GHz for \dcopns(3-2); 216.94560~GHz for
CH$_3$OH(5(1,4)-4(2,2));
215.59595~GHz for SiO$_{v=1}$(5-4); and 217.23853~GHz for
DCN(3-2). Each of these windows has a total bandwidth of 58 MHz ($\sim$ 81 \kmsns)  
and a 0.2 \kms velocity resolution.
 The data are Hanning smoothed.

 We performed the standard cleaning procedure in CASA using
the task {\it clean}. For the continuum and CO data, we used
the Briggs weighting with a robust number of 0.5 (restoring beam $\sim$ 0.25\arcsec). 
To have the best sensitivity, we used the natural weighting for the line
data, including \ntdpns(3-2), \dcopns(3-2), \ceions(3-2), and DCN(3-2)
(restoring beam $\sim$ 0.3\arcsec).
For the compact data, we performed three
iterations of phase-only self-calibration, which improved the sensitivity
by a factor of two. 
The self-calibration solutions to the continuum image were applied to
other spectral windows. The final continuum sensitivity achieved is
0.12 mJy beam$^{-1}$ for the compact-configuration data; 0.042 mJy
beam$^{-1}$ for the extended-configuration data; 0.040 mJy beam$^{-1}$
for the combined data.  The line sensitivity for the compact
configuration data is $\sim10\:$mJy beam$^{-1}$ per 0.1~\kmsns; for
the extended configuration data is $\sim8\:$mJy beam$^{-1}$ per
0.1~\kmsns; for the combined data is $\sim6\:$mJy beam$^{-1}$ per
0.1~\kmsns.
 In the following, all analysis
are based on the combined images.

\subsection{VLA Observations}

The Karl G. Jansky {\it Very Large Array (VLA)} of NRAO was pointed
towards C1-S in its D configuration on 2014 August 12 and September
11, with the same correlator setup to observe NH$_3$ (J,K) = (1,1) and
(2,2) lines. We used a bandwidth of 8 MHz with a channel width of 7.8
kHz (corresponding to $\sim$0.1 \kmsns) in dual polarization. The
antenna gain, bandpass, and flux variations were calibrated by
standard calibrators J1851+0035, J1743-0350, and 3C286, respectively,
for both observing sessions. Data reduction was performed in CASA
4.2.2. The calibrated visibilities from the two sessions were imaged
together. During imaging, we smoothed the channel width to 0.2 \kms
and used a natural weighting, in order to optimize the signal-to-noise
ratio.  The resulting data cubes have an rms noise of 4 mJy per
3\arcsec\ synthesized beam.

We will use these NH$_3$ data to derive gas kinetic temperature
following the routines outlined in
\citet{2012ApJ...745L..30W,2014MNRAS.439.3275W}, which use the
detailed method developed by \citet{1983ARA&A..21..239H} and
\citet{2008ApJS..175..509R}. We model the NH$_3$ (J,K) = (1,1) and
(2,2) spectra simultaneously with five free parameters, including
kinetic temperature, excitation temperature, NH$_3$ column density,
velocity dispersion, and line central velocity.  The model fitting is
performed on pixels with $>6 \sigma$ NH$_3$ (J,K) = (1,1) integrated
intensity. These temperature measurements are discussed below in
section \ref{subsec:cont}, specifically for their use in providing
partial constraints on the temperatures needed for mass estimates from
the mm dust continuum emission, assuming coupling of gas and dust
temperatures.

\section{Results}\label{S:results}

\subsection{1.3~mm Dust Continuum Emission and $\rm NH_3$-Derived Temperature Map}\label{subsec:cont}

\begin{figure*}[htb!]
\epsscale{1.2}
\plotone{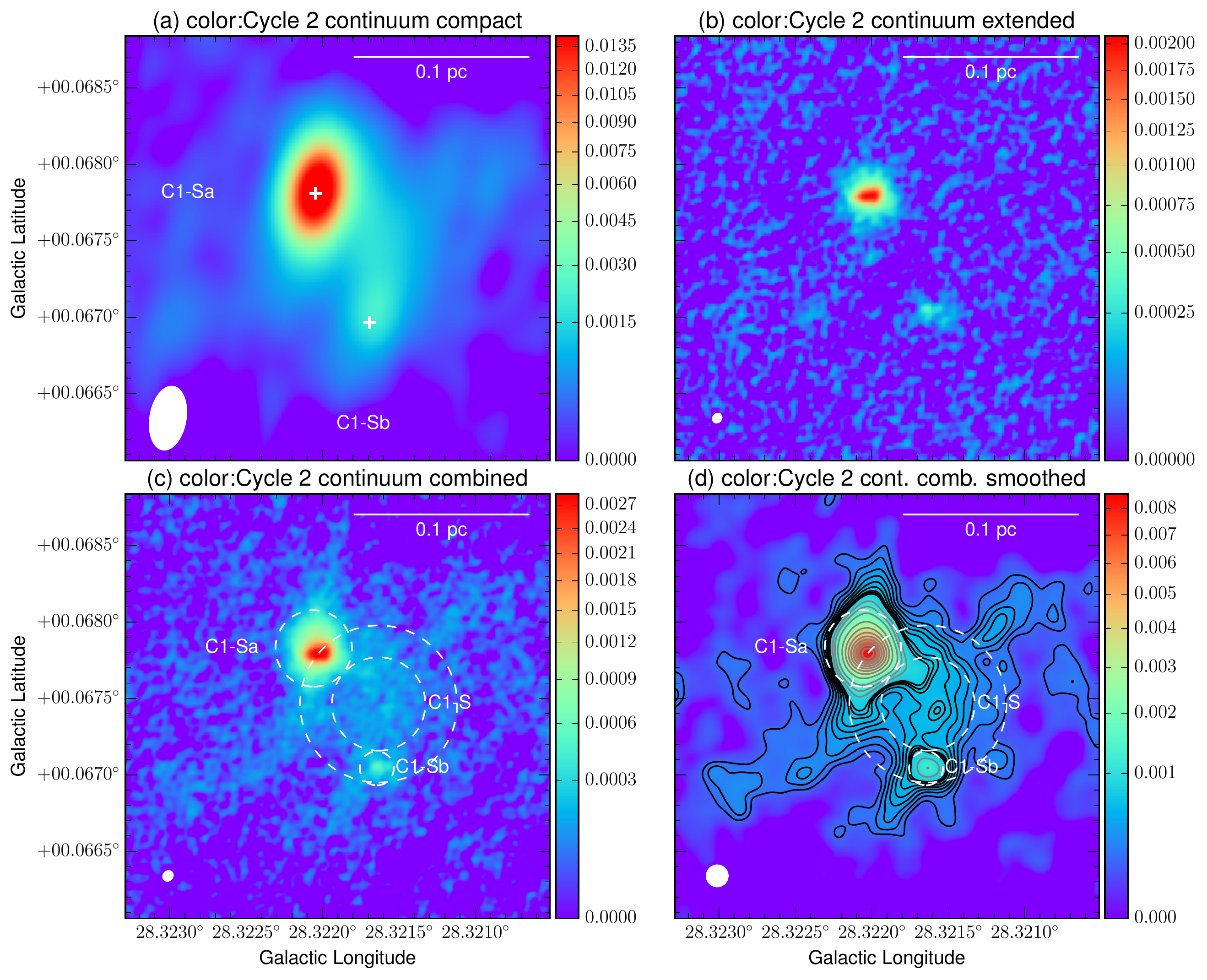}
\caption{
1.30~mm continuum images of C1-S region: {\it (a) Top left:}
compact-configuration (1$\sigma$ rms 0.12 mJy beam$^{-1}$); {\it (b)
  Top right:} extended-configuration (1$\sigma$ rms 0.042 mJy
beam$^{-1}$); {\it (c) Bottom left:} combined (1$\sigma$ rms 0.040 mJy
beam$^{-1}$); {\it (d) Bottom right:} combined, smoothed to 0.5\arcsec
(1$\sigma$ rms 0.080 mJy beam$^{-1}$). 
The black contours show 2, 3, 4, 5, 6, 7, 8, 8.5, 9, 10$\sigma$.
The gray contours show 12, 15, 20, 30, 40, 50...$\sigma$.
The color scale bars have units
of \jybns. The synthesized beams are shown as white filled ellipses in
the lower-left corners of each panel. 
The white plus signs in panel
(a) show the protostellar sources, C1-Sa and C1-Sb, identified by
T16. The white dashed circles in panels (c) and (d) show the apertures
we use later in the paper to measure fluxes from C1-Sa, C1-Sb and the
inner and outer regions of the C1-S candidate massive pre-stellar
core.
\label{fig:cont1}}
\end{figure*}

Figure \ref{fig:cont1} shows the 1.30~mm continuum images of the C1-S
region. Note, that T16 already presented compact configuration images
of the entire observed region and in this paper we will focus mostly
on the sub-region shown in Figure~\ref{fig:cont1}.  Panel (a) shows
the compact-array data with 1.5\arcsec~beam.  The two protostellar
cores C1-Sa and C1-Sb from T16 are labeled with plus signs. Panel (b)
presents the extended-array data with a beam size of 0.2\arcsec.
Panel (c) shows the combined data with a final resolution also of
0.2\arcsec. The synthesized beam corresponds to a physical scale of
about 0.005~pc, i.e., $\sim$1000~AU, given the adopted distance to the
source of 5.0 kpc. Panel (d) shows the same data as (c), but now
smoothed to about 0.5\arcsec\ resolution.
We also overlay contours for the continuum emission in Panel (d).
There is a small continuum peak in C1-S.

We define C1-Sa and C1-Sb with two circular regions (the dashed
circles in Figures \ref{fig:cont1}c and d).  The C1-Sa core shows a
relatively smooth profile as a function of radius from the center,
resolved with $\gtrsim$ 10 beams and with no sign of fragmentation.
The bright central region shows elongation in the direction of
Galactic longitude.  The elongation is resolved with 3 synthesized
beams.  In the continuum image of Figure \ref{fig:cont1}c, a bipolar
structure is seen stretching out from C1-Sa in the approximately   
(Galactic) north-south directions. This structure coincides with
the CO outflow axis (T16, and see, below, \S\ref{subsec:co}),
indicating that there is dust in this outflow or along its cavity
walls that is likely heated to temperatures greater than those of dust
in the ambient environment.

Our defined core region of C1-Sb is much smaller in size compared to
C1-Sa, with only $\sim4$ beams across its diameter. The peak flux
density is also several times smaller than that of C1-Sa.

The angular distance between C1-Sa and C1-Sb is $\sim3$\arcsec,
corresponding to about 15,000~AU (0.073~pc). Interestingly, in this
region between C1-Sa and C1-Sb, we see a diffuse distribution of
1.30~mm dust emission. At least in projection, it appears to connect
C1-Sa and C1-Sb with an arc-like shape.  It does not show any clear
centrally peaked profile. This structure spatially overlaps with the
\ntdpns(3-2) core C1-S defined in T13.  In the following, we show the
new high-resolution \ntdp data and we make new definitions of the size
of this core, i.e., with an inner and outer scale considered, which
are also shown in Figures~\ref{fig:cont1}c and d.

Our primary method of estimating masses is based on 1.3~mm dust
continuum emission, following eq. (7) of T13, i.e., using the
opacities of the moderately coagulated thin ice mantle model of
\citet{1994A&A...291..943O}.  The mass surface density from mm
continuum is:
\begin{equation}\label{eq:Sigmamm}
\begin{split}
\Sigma_{\rm mm} = & 5.53\times
10^{-3} \left(\frac{S_\lambda/\Omega}{{\rm MJy/sr}}\right) \left(\frac{\kappa_\lambda}
{0.01\:{\rm cm^2\:g^{-1}}}\right)^{-1} \lambda_{1.30}^3 \\
& \times \left[{\rm exp}\left(1.106 T_{d,10}^{-1}
\lambda_{1.30}^{-1}\right)-1\right]\:{\rm g\:cm^{-2}}
\end{split}
\end{equation}
where $\lambda_{1.30}=\lambda/1.30\:{\rm mm}$ and
$T_{d,10}=T_d/10\:{\rm K}$. We choose $\kappa_\nu = 5.95\times
10^{-3}\:{\rm cm^2\:g^{-1}}$ and assume 30\% uncertainties in
  these opacities.  This method of mass estimation depends on the
adopted dust temperature, which is somewhat uncertain. We thus
calculate core masses for a range of temperatures.

\begin{figure*}[htb!]
\epsscale{1.}
\plotone{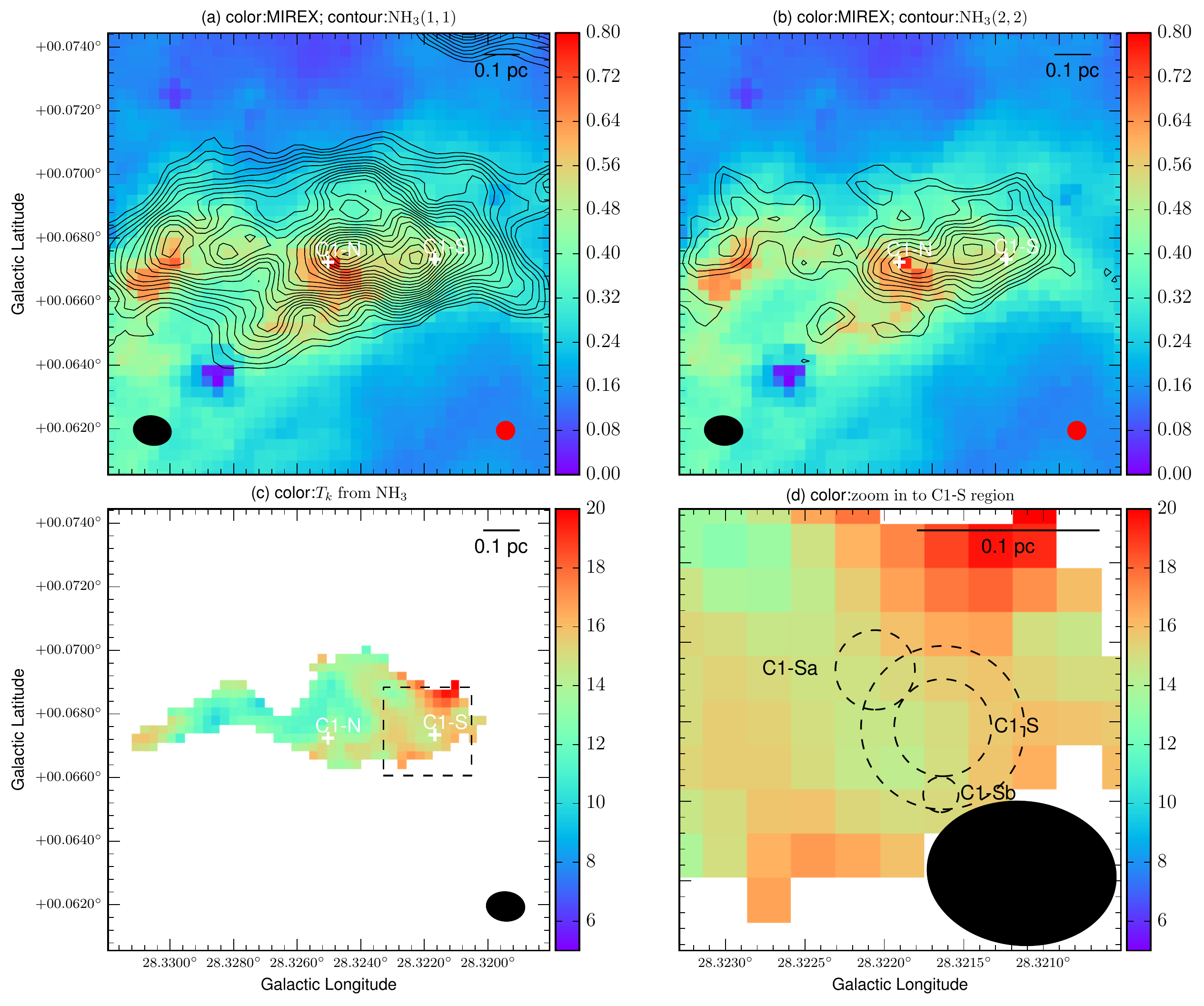}
\caption{
(a) Zeroth moment map of NH$_3$(1,1) integrated from 76.9 to 81.9 \kms 
  (contours, shown from 3, 4, 5$\sigma$, etc., with $\sigma$ being 4 mJy km s$^{-1}$). 
  The integration covers the main hyperfine structure. 
  The background image shows
  the MIR+NIR extinction map (in $\rm g\:cm^{-2}$) from \citet{2014ApJ...782L..30B}.
 The white plus signs represent the C1-N and C1-S cores from T13.
  The red circle in the lower-right shows the resolution beam of the
  extinction map.  The black ellipse in the lower-left shows the
  synthesized beam of the {\it VLA} observations.
(b) As panel (a), but now showing zeroth moment map of NH$_3$(2,2).
(c) $T_k$ map derived from the {\it VLA} NH$_3$ observations. The
color scale bar has the unit of K. The filled ellipse is the beam of
the {\it VLA} observations. The dashed inset shows the zoom-in region
of C1-S in panel (d).
(d) Zoom-in view to the C1-S region in the $T_k$ map. The black dashed
circles represent the C1-S, C1-Sa and C1-Sb core definitions (see
\S\ref{subsec:cont} and \S\ref{subsubsec:n2dpcore}).  The filled
ellipse in the lower-right is the beam of the {\it VLA} observations.
\label{fig:tkin}}
\end{figure*}

Figure~\ref{fig:tkin} shows the gas kinetic temperature $T_k$ measured
from the {\it VLA} NH$_3$ observations. A temperature of
$T_k\simeq13\:$K to $16\:$K is observed towards the C1-S region, with
slightly warmer temperatures just to the north.  However, given the
relatively low critical density ($\sim10^3\:$cm$^{-3}$) of the NH$_3$
transitions (i.e., compared to $\sim 3\times10^6\:{\rm cm}^{-3}$ for
\ntdpns(3-2)), 
we expect that the temperature traced by NH$_3$
is likely influenced by the lower density envelope
material around the C1-S \ntdp core .  The line width of the NH$_3$
lines are also broader than \ntdpns(3-2), discussed below. The
protostellar cores are not well resolved by the NH$_3$ observations,
so the ability to see any warmer gas associated with these
protostellar envelopes is compromised by beam dilution.

T13 estimated $T_{d}= 10\pm3$~K, based on the region appearing dark at
up to $\sim100\:{\rm \mu m}$.  Thus for the C1-S starless core,
defined in the next subsection, we will adopt $T_{\rm
  dust}=10\pm3$~K. In the models of \citet{2014ApJ...788..166Z} for
high-mass protostellar cores and those of \citet{2015ApJ...802L..15Z}
for low-mass cores, the envelope gas temperature is between $\sim$15~K
to 30~K during the phases of evolution relevant to C1-Sa (i.e., early
phases, given the expectation of it being a massive protostellar core,
see below) and to C1-Sb (i.e., most evolutionary phases relevant to
low-mass star formation, given the expectation it is a lower-mass
protostar, see below).  Here we adopt a fiducial temperature of 20 K
for the protostellar cores C1-Sa and C1-Sb, with lower and upper
bounds of 15 K and 30 K, respectively.  
The measured core flux density in C1-Sa is 40.9 mJy and in
C1-Sb is 2.31 mJy, with systematic uncertainties in these absolute 
values expected to be $\lesssim 10\%$ so that overall mass uncertainties 
are dominated by other factors, such as temperature and dust opacity 
uncertainties. Table \ref{tab:mass} shows the
resulting core masses based on the observed mm continuum fluxes, given
these temperature uncertainties.
The 1$\sigma$ continuum mass sensitivity is 0.03 $M_\odot$ per beam
at $T$ = 20~K.

The mass estimates of the C1-S starless core are discussed below in
the next subsection. The protostellar envelope of C1-Sa has an
estimated mass of about $30\:M_\odot$, with a range of values from
$\sim20$ to $50\:M_\odot$ given the temperature uncertainties. Based
on the momentum flux of its CO outflow, T16 estimated the mass of the
C1-Sa protostar to be $\lesssim 3\:M_\odot$. Thus C1-Sa is consistent
with being a massive protostellar core that is in a relatively early
phase of collapse. We find that C1-Sb has about a 14 times smaller mm
continuum flux and thus the mass estimates for its envelope are just a
few Solar masses. Given that its CO outflow momentum flux is only
moderately smaller than that of C1-Sa, this suggests that C1-Sb is a
low or intermediate mass protostar at a more advanced evolutionary
stage of its collapse. This is also consistent with C1-Sb's larger
outflow cavity opening angle (T16, see also \S\ref{subsec:co}).

\begin{table*}
\centering
\begin{threeparttable}
\small
\caption{Core Masses}\label{tab:mass}
\begin{tabular}{ccccccc}
\hline {{Protostellar Cores}} & {{$l$}} & {{$b$}} & {{$R_{\rm c}$}} & {{$M_{\rm c,mm}$ ($T=15\:$K)}} & {{$M_{\rm c,mm}$ ($T=20\:$K)}} & {{$M_{\rm c,mm}$ ($T=30\:$K)}} \\
 & {{(deg)}} & {{(deg)}} & {{(0.01~pc)}} & {{($M_\odot$)}} & {{($M_\odot$)}} & {{($M_\odot$)}}\\
\hline
C1-Sa & 28.32209 & 0.06770 & 2.18 & 49.5  & 33.5$^a$   & 20.2  \\ 
C1-Sb & 28.32168 & 0.06692 & 0.97 & 3.44  & 2.33$^a$   & 1.41  \\
\hline 
{{Starless Core}} & {{$l^d$}} & {{$b^d$}} & {{$R_{\rm c}$}} & {{$M_{\rm c,mm}$ ($T=7\:$K)}} & {{$M_{\rm c,mm}$ ($T=10\:$K)}} & {{$M_{\rm c,mm}$ ($T=13\:$K)}}\\
 & {{(deg)}} & {{(deg)}} & {{(0.01~pc)}} & {{($M_\odot$)}} & {{($M_\odot$)}} & {{($M_\odot$)}}\\
\hline
C1-S inner (tot)$^b$ & 28.32167 & 0.06734 & 2.67 & 53.5 & 25.5$^a$  & 11.6 \\ 
C1-S inner$^c$ & ... & ... & ... & 43.7 & 17.1$^a$  & 4.97 \\ 
\hline
C1-S outer (tot)$^b$ & 28.32167 & 0.06734 & 4.48 & 123  & 58.8$^a$  & 26.8\\ 
C1-S outer$^c$ & ... & ... & ... & 112  & 48.8$^a$  & 19.0\\
\hline
\end{tabular}
\begin{tablenotes}
\small
\item $^a$ Fiducial case.
\item $^b$ Masses estimated using total mm continuum fluxes.
\item $^c$ Masses estimated using background subtracted mm continuum fluxes.
\item $^d$ Defined at the center of the circular core. 
\end{tablenotes}
\end{threeparttable}
\end{table*}

\subsection{\ntdpns(3-2) Emission}

\subsubsection{Definition of the C1-S Pre-stellar Core}\label{subsubsec:n2dpcore}

\begin{figure*}[htb!]
\epsscale{1.2}
\plotone{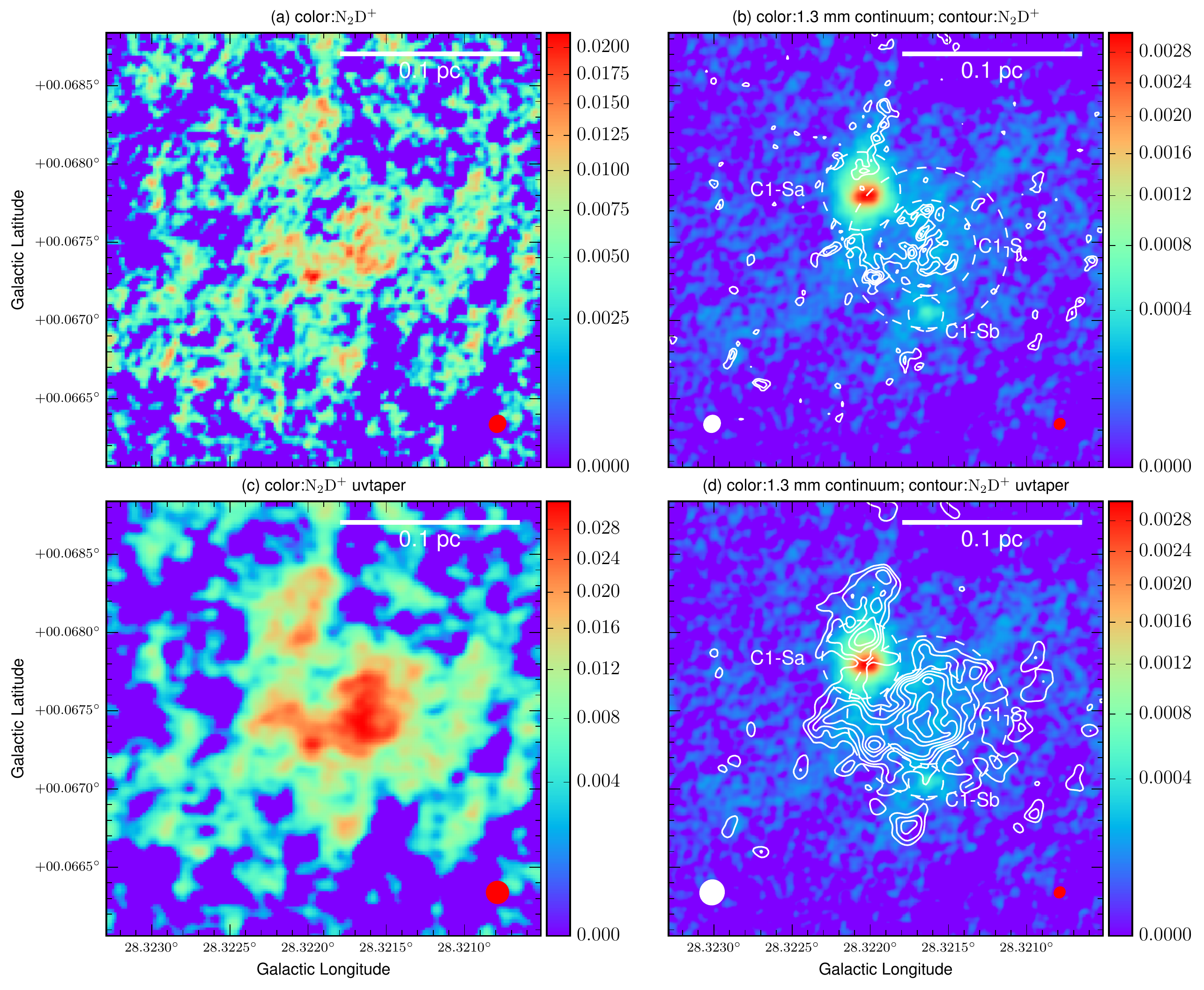}
\caption{
\ntdp observations: (a) Combined \ntdpns(3-2) 0th-moment map. The
integration is from 76.9 to 81.9 \kms (i.e., centered on 79.4 \kms
velocity measured by T13). The color scale bar has a unit of \jyb
\kmsns. The synthesized beam ($\sim$ 0.3\arcsec) is shown as a red
filled ellipse at lower-right.
(b) \ntdpns(3-2) 0th-moment contours overlaid on the 1.3 mm continuum
image. The continuum image and color scheme is the same as
Figure \ref{fig:cont1}c. The \ntdpns(3-2) contours start from 3$\sigma$
and increase in step of 1$\sigma$ 
($\rm \sigma=3~mJy~km~s^{-1}$). 
The red filled ellipse at
lower-right is the synthesized beam for the continuum image. The white
filled ellipse at lower-left is the beam for the \ntdpns(3-2) contour
map. The dashed circles are the core definitions for C1-Sa, C1-Sb,
``C1-S inner'' and ``C1-S outer.''
(c) Same as (a), but now the \ntdpns(3-2) 0th-moment map
is uv-tapered to $\sim$  0.5\arcsec\ resolution.
(d) Same as (b), but now the \ntdpns(3-2) 0th-moment contour map 
is uv-tapered to $\sim$  0.5\arcsec\ resolution.
\label{fig:ntdp1}}
\end{figure*}

Figure \ref{fig:ntdp1} shows the \ntdpns(3-2) 0th-moment map of the
C1-S region, integrated from 76.9 to 81.9 \kms (i.e., centered on the
79.4 \kms measured for the C1-S core by T13). Several versions of the
map are shown. In panel (a), we show the combined data of \ntdpns(3-2)
with a final synthesized beam size of 0.3\arcsec. In panel (b), the
same \ntdp image is overlaid as contours on top of the 1.30 mm
continuum image from Figure \ref{fig:cont1}c. Since noise
  features are quite visible across the image in panel (a), in panel
(c) we show a smoothed image, obtained by applying an outer uv-taper
to the \ntdp data to have a
$0.57\arcsec\times0.54\arcsec~P.A.=89.59\arcdeg$ beam. Then in panel
(d), we show the smoothed \ntdp image on top of the continuum image.
One can see from Figure~\ref{fig:ntdp1}, especially panels (c)
  and (d), that there is centrally concentrated \ntdp emission
inbetween C1-Sa and C1-Sb. As we discuss below, this structure, which
we will refer to as a ``core'', contains a large fraction of the
\ntdpns(3-2) flux of the C1-S core defined by T13. We remind that
the \ntdpns(3-2) flux of C1-S is the strongest source of such
emission in the wider region and is highly localized in position and
velocity space, i.e., to $\sim0.1$pc spatial and $\lesssim 0.5\:{\rm
  km\:s}^{-1}$ velocity scales. In addition to the radial gradient
in \ntdpns(3-2) intensity, C1-S also shows some internal
substructure in this emission at the factor of about two to three
level. Such variation is broadly consistent with the astrochemical
models of \citet{2016ApJ...833..274G} of a monolithic core that is
partially supported by turbulence. In these models the abundance of
\ntdp increases by several orders of magnitude and depends on
local density, temperature and other gas properties, so fluctuations
of a factor of a few are readily explained as being in response to
local transient turbulent fluctuations, without necessarily implying
that these structures will fragment into separate entitites.

We define ``C1-S inner'' as the circle centered approximately on the
peak \ntdpns(3-2) 0th-moment contours, while extending out as far as
possible to just reach, in projection, the C1-Sa and C1-Sb
protostellar cores. C1-S inner contains 23\% of the total
  \ntdpns(3-2) flux  in the high-resolution image (panel b). 
We also define ``C1-S outer'' with a circular aperture that shares the
same center but expanded to cover a larger fraction (44\%)  of
the \ntdpns(3-2) flux in the region  (see Figure \ref{fig:ntdp1}b
and d). C1-S outer partially overlaps C1-Sa and fully overlaps C1-Sb.
In any later analysis, pixels in C1-Sa and C1-Sb, e.g., containing mm
continuum flux or \ntdpns(3-2) flux, are excluded from contributing to
C1-S outer.

\begin{figure*}[htb!]
\epsscale{1.}  
\plotone{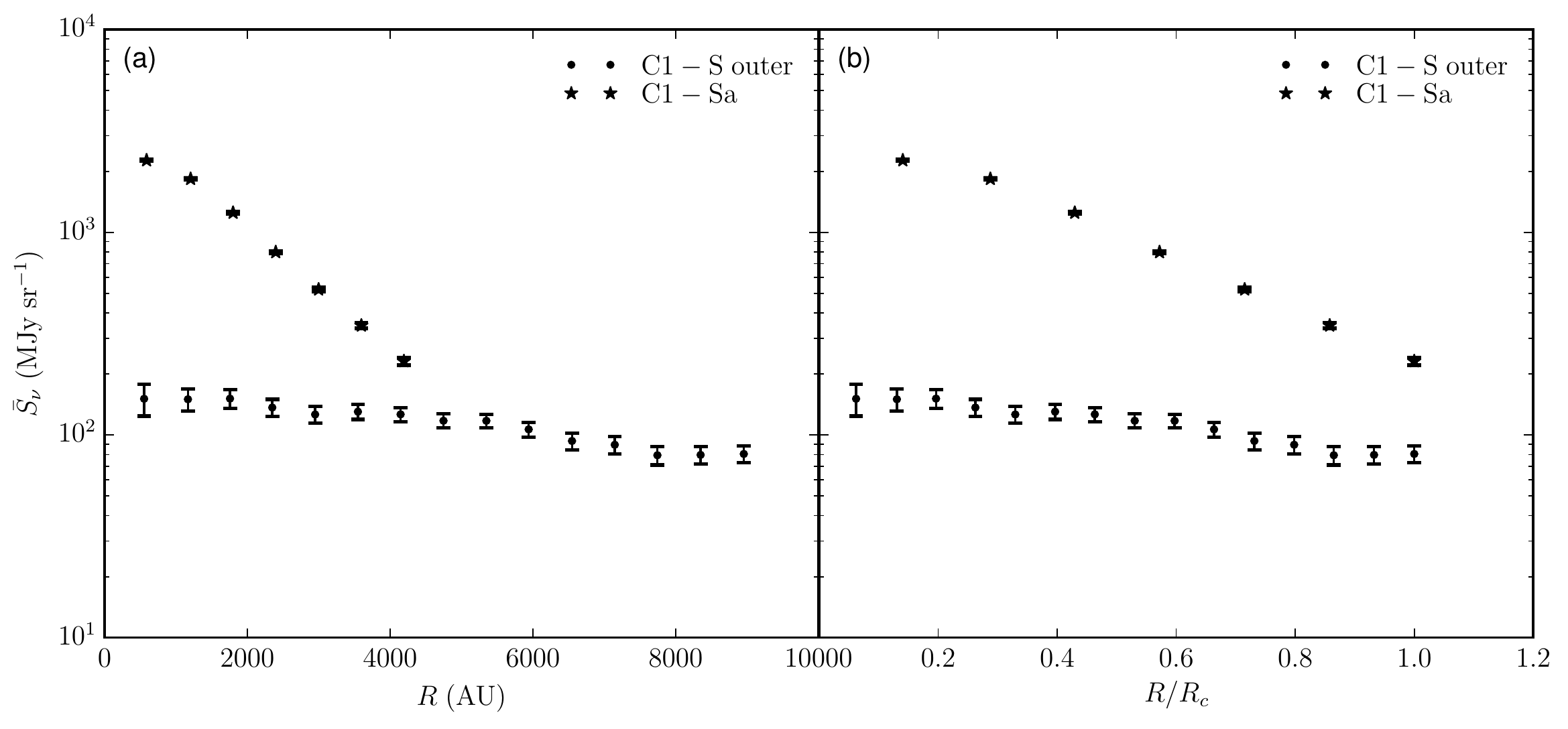}
\caption{
(a) 1.3 mm continuum flux density $\bar{S}_{\rm 1.3mm}$ as a function
  of projected radius, $R$, for C1-S outer (dots) and C1-Sa (stars).
  The measurement is averaged within concentric annuli that have
  widths equal to the beam size.  The error bars show uncertainties
  due to observational noise (image rms noise normalized by the 
  square root number of independent beams inside the annuli).
(b) Same as panel (a), but the radii are normalized to the core radii. 
\label{fig:snuradial}}
\end{figure*}

Note here that we are defining C1-S inner and outer by their
centrally-peaked \ntdp emission, unlike C1-Sa or C1-Sb that were
defined by centrally-peaked continuum emission. As shown in Figure
\ref{fig:snuradial}, C1-Sa shows a centrally-peaked 1.3~mm continuum
intensity profile, while C1-S's profile is much shallower. The lack of
a centrally peaked continuum morphology in C1-S, compared to C1-Sa and
C1-Sb, is thus evidence for it being a starless, pre-stellar core,
rather than a protostellar core. Note also that the extent of material
that is dynamically associated with C1-S, e.g., bound to it, may be
larger than the example aperture of C1-S outer. Fig. \ref{fig:ntdp1}c
shows this more extended \ntdpns(3-2) emission, although a small part
of this may be associated with the C1-Sa and C1-Sb protostars.

\begin{table*}
\centering
\begin{threeparttable}
\caption{Continuum flux measurement data for Figures \ref{fig:snuradial} and \ref{fig:massradial}.}\label{tab:measure}
\begin{tabular}{ccccc}
\hline {{$R_{\rm C1-Sa}$}} & {{$\overline{F}_{\rm C1-Sa}(R)$}} & {{$R_{\rm C1-S~outer}$}} & {{$\overline{F}_{\rm C1-S~outer}(R)$}} & {{$\overline{F}_{\rm C1-S~outer~sub}(R)$}} \\
{{AU}} & {{MJy sr$^{-1}$}} & {{AU}} & {{MJy sr$^{-1}$}} & {{MJy sr$^{-1}$}} \\
\hline
588  & 2266(25)& 554.5  & 150.9(27.0) & 117.7(36.4) \\
1205 & 1833(19)& 1168.3 & 150.1(19.0) & 116.9(30.9) \\
1800 & 1249(15)& 1757.5 & 151.2(15.8) & 118.0(29.0) \\
2397 & 798(13) & 2349.2 & 136.6(13.4) & 103.4(27.7) \\
2998 & 522(11) & 2949.7 & 126.3(12.0) & 93.1(27.1)  \\
3597 & 346(10) & 3549.2 & 130.3(10.9) & 97.2(26.6)  \\
4193 & 231(10) & 4146.6 & 126.4(10.1) & 93.3(26.3)  \\
 -   & -       & 4747.3 & 117.6(9.4)  & 84.4(26.1)  \\
 -   & -       & 5347.9 & 117.5(8.9)  & 84.4(25.9)  \\
 -   & -       & 5940.6 & 106.6(8.9)  & 73.4(25.9)  \\
 -   & -       & 6547.4 & 93.4(8.8)   & 60.3(25.9)  \\
 -   & -       & 7146.4 & 89.5(8.7)   & 56.4(25.8)  \\
 -   & -       & 7743.3 & 79.5(8.3)   & 46.3(25.7)  \\
 -   & -       & 8348.5 & 79.8(8.0)   & 46.6(25.6)  \\
 -   & -       & 8953.4 & 80.6(7.7)   & 47.5(25.5)  \\
\hline
\hline
\end{tabular}
\end{threeparttable}
\end{table*}

The 1.3~mm continuum-derived  mass estimates for C1-S inner and
outer shown in Table~\ref{tab:mass} range from about $25$ to
$60\:M_\odot$ for the fiducial choice of assumptions, i.e., a dust
temperature of 10~K. Temperature uncertainties from 7 to 13~K induce
about a factor of two uncertainty in the mass estimates. Note,
however, that these mass estimates have so far been based on the total
flux observed towards a given aperture. Clump envelope substraction,
 i.e., where the average flux density in an annular region
  extending to twice the core radius (and excluding the protostars
  C1-Sa and C1-Sb) is subtracted (see \S\ref{subsubsec:dynamics}), 
leads to modest reductions in these
mass estimates by factors of about 0.64 for C1-S inner and 0.83 for
C1-S outer in the fiducial case. Thus, the mass that is associated
with the observed \ntdpns(3-2) emission, especially when considered at
its outer scale, appears to be of sufficient quantity to form a
massive star, if this core were to undergo quasi-monolithic collapse.

C1-S is a highly deuterated core. The deuteration analysis of
\citet{2016ApJ...821...94K}, which was applied to the 3.5\arcsec
radius C1-S aperture of T13 found $D_{\rm frac}^{\rm N2H+}\equiv [{\rm
    N_2D^+}]/[{\rm N_2H^+}] \simeq0.2$ to 0.7. We thus expect the
actual deuteration level in C1-S outer and inner to be even
higher. Such high levels of deuteration are expected to be achieved in
gas that is kept in a cold ($T\lesssim 20\:$K) and dense state for a
relatively long time, i.e., compared to the local free-fall time. Such
conditions are found to be common in low-mass pre-stellar cores
\citep{2012A&ARv..20...56C} and may be expected to arise in more
massive pre-stellar cores also, assuming such cores exist.

In summary, C1-S shows no obvious sign of star formation. It has no
centrally peaked continuum profile. It has no corresponding CO
outflows (as will be shown below in \S\ref{subsec:co}). In {\it
  Herschel} 70 $\micron$ and 100 $\micron$ images, this entire region
has no sign of point sources (\citealt{2016ApJ...829L..19L}; Lim et
al., in prep.). While C1-Sa and C1-Sb have some level of \ntdpns(3-2)
emission, this tracer does not show obvious concentrations at the
locations of their continuum emission peaks.  Meanwhile, C1-S has a
strongly concentrated \ntdpns(3-2) structure away from the
protostars. We thus conclude that C1-S is likely to be a highly
deuterated massive starless core.

\subsubsection{Core Structure}\label{subsec:structure}

\begin{figure*}[htb!]
\epsscale{1.1}
\plotone{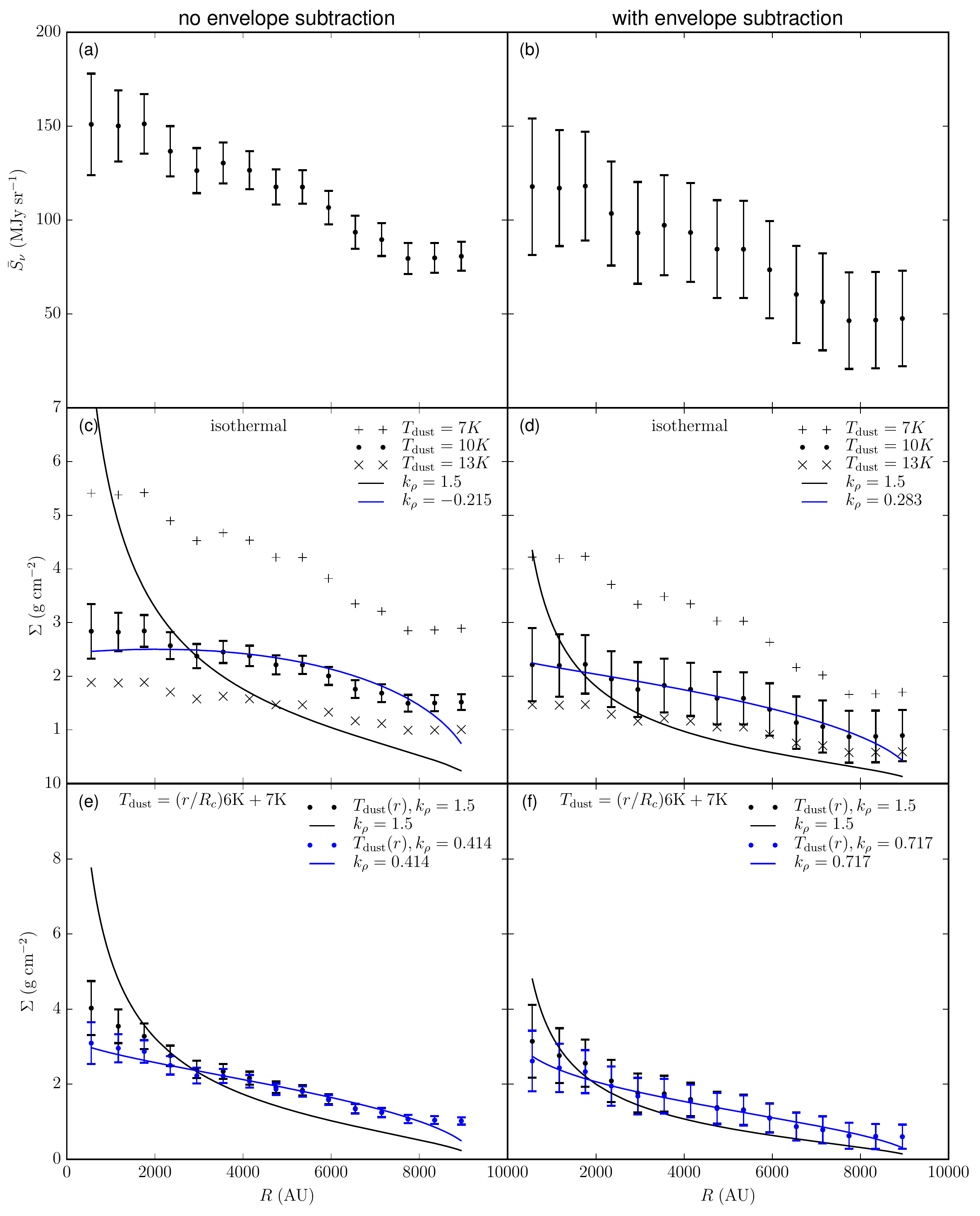}
\caption{
(a) 1.3 mm continuum flux density $\bar{S}_{\rm 1.3mm}$ as a function
  of projected radius, $R$, for C1-S outer. The measurement is averaged
  within concentric annuli that have widths equal to the beam size.
  The error bars show uncertainties due to observational noise.
(b) Same as panel (a), but after subtraction of the clump envelope, as
  evaluated from an annulus from $R_c$ to $2R_c$. 
(c) Mass surface density,
$\Sigma$, as a function of projected radius $R$ for C1-S outer
assuming isothermal conditions. The fiducial dust temperature is
$10\:$K (black dots). Also shown are cases with $7\:$K and $13\:$K.
The solid lines show $\Sigma$ profiles of model cores with power-law
volume density profiles $\rho~\propto~r^{-k_\rho}$, with best-fit case
of $k_\rho=-0.215$ shown with a blue line and fiducial reference case
of $k_\rho=1.5$ shown with a black line. The normalization of these
profiles is based on a $\chi^2$ minimization to the data.  (d) Same as
panel (c), but with clump envelope subtracted.  The best-fit
$k_\rho=0.283$.
(e) A linear temperature profile with 3D radius from 7 K (at core
center) to 13 K (at core surface) is adopted for the mass surface
density estimation. Two $\Sigma$ profiles are shown for the fiducial 
MT03 turbulent core ($k_\rho=1.5$) and for the best-fit value of $k_\rho$.
The black curve is the model for
$k_\rho=1.5$, and the blue curve is the model involving $k_\rho$,
which gives the best-fit $k_\rho=0.414$.
(f) Same as panel (e), but with clump envelope subtracted.
The best-fit $k_\rho=0.717$.
\label{fig:massradial}}
\end{figure*}

Figure \ref{fig:massradial} shows the radial structure of C1-S outer
as inferred from thermal dust continuum emission.  In panel (a) we
show the total $\bar{S}_{\rm 1.3mm}(R)$ (averaged in annuli), while in
(b) we show the result after envelope subtraction. This background
envelope level is assessed as the average value of $\bar{S}_{\rm
  1.3mm}$ in an annulus from one to two core radii (excluding the
C1-Sa and C1-Sb regions). We estimate the uncertainty associated with
this process as equal to the level of fluctuations seen in 8
independent regions of the annulus, i.e., that have linear dimensions
comparable to the radius of the core. The standard deviation of these
mean values is adopted as the uncertainty, which is about 70\% of
overall envelope value.  In panel (c), we calculate the mass surface
  density profile from the total 1.3~mm flux density based on
  Eq.\ref{eq:Sigmamm}.  Since C1-S is a candidate starless core, we
  first assume $T_{\rm dust}=10\:$K across the core (black dots).  We
  also calculate two more cases where $T_{\rm dust}=7\:$K and $T_{\rm
    dust}=13\:$K, which we use to assess the uncertainty in the mass
  surface density estimates.

Following the turbulent core model of MT03, we assume a core volume
density profile following a singular polytropic spherical structure:
$\rho~\propto~r^{-k_\rho}$, where $k_\rho=1.5$ in the fiducial case.
We project the volume density to surface density $\Sigma$ by
integrating along the line of sight.  Then we find the best fitting
model, first for the case of $k_\rho=1.5$ (black line in panel c). It
has a density at core surface (9250 AU) (i.e., the radius of C1-S
  outer)  of $n_H=8.40\times10^{5}\:{\rm cm}^{-3}$. However, this
model is not a very good fit to the data.  We then allow $k_\rho$ to
vary, which yields a best-fit model with $k_\rho=-0.215$ (blue line),
i.e., $\rho$ increasing towards the core surface, which is not
expected for physically realistic models.  We repeat the analysis for
the clump envelope-subtracted profiles shown in panel (d). Now the
best-fit $k_\rho$ is 0.283. This has density declining with radius,
but is still much shallower than the fiducial case of MT03.

In panels (e) and (f), we also investigate models with a temperature
gradient in the core.  As an example, we consider a linear gradient
with $T=7K$ at the core center and $T=13K$ at the core surface,
similar to the range of values seen in some lower-mass pre-stellar
cores \citep{2007A&A...470..221C}.  The newly-derived $\Sigma$ profile
are shown as dots.  One can see $\Sigma$ is more centrally-peaked
compared to the isothermal case.  Again we model the core and the
best-fit $k_\rho=0.414$ for the total flux case and $k_\rho=0.717$ for
the envelope-subtracted case. Opacity variations within the core,
e.g., due to grain growth in the denser central regions, could also
lead to changes in the derived density structures. Lacking any good
constraints on temperature and opacity variations within the core, we
conclude that it is difficult to determine the core structure from the
current data. We also note that as a starless core contracts to
becoming a pre-stellar core on the verge of forming a protostar, its
density profile is expected to steepen \citep{1977ApJ...214..488S}.

\subsubsection{Core Kinematics}\label{subsubsection:ck}

\begin{figure*}[htb!]
\epsscale{.35}
\plotone{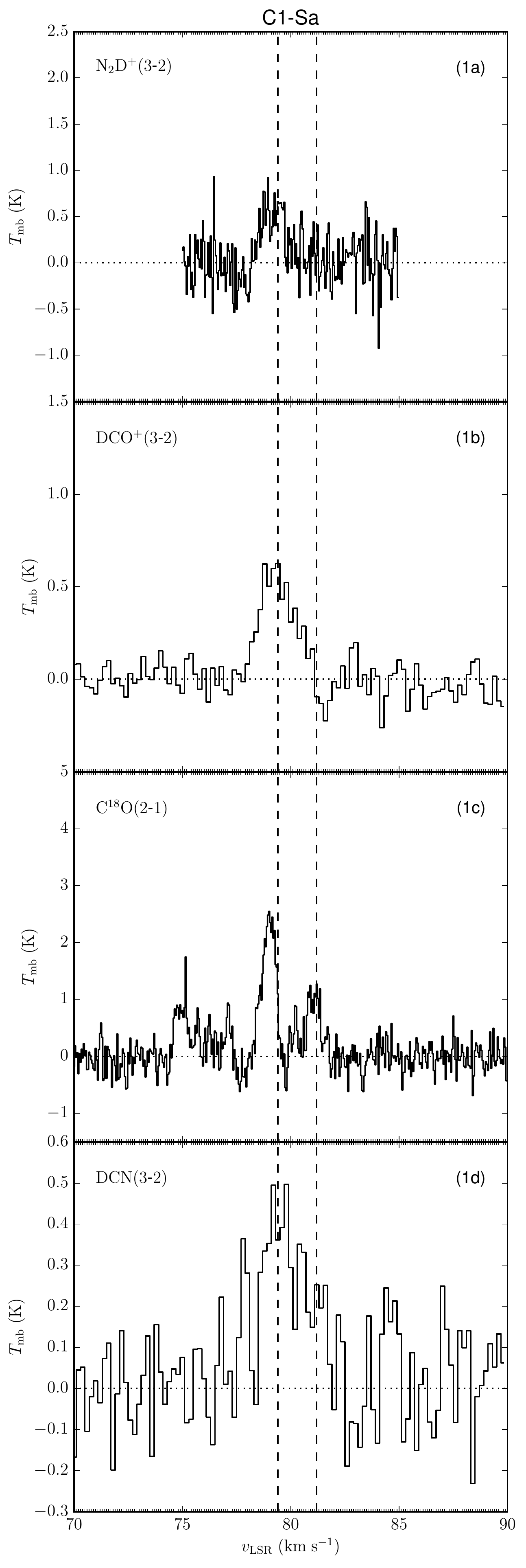}
\plotone{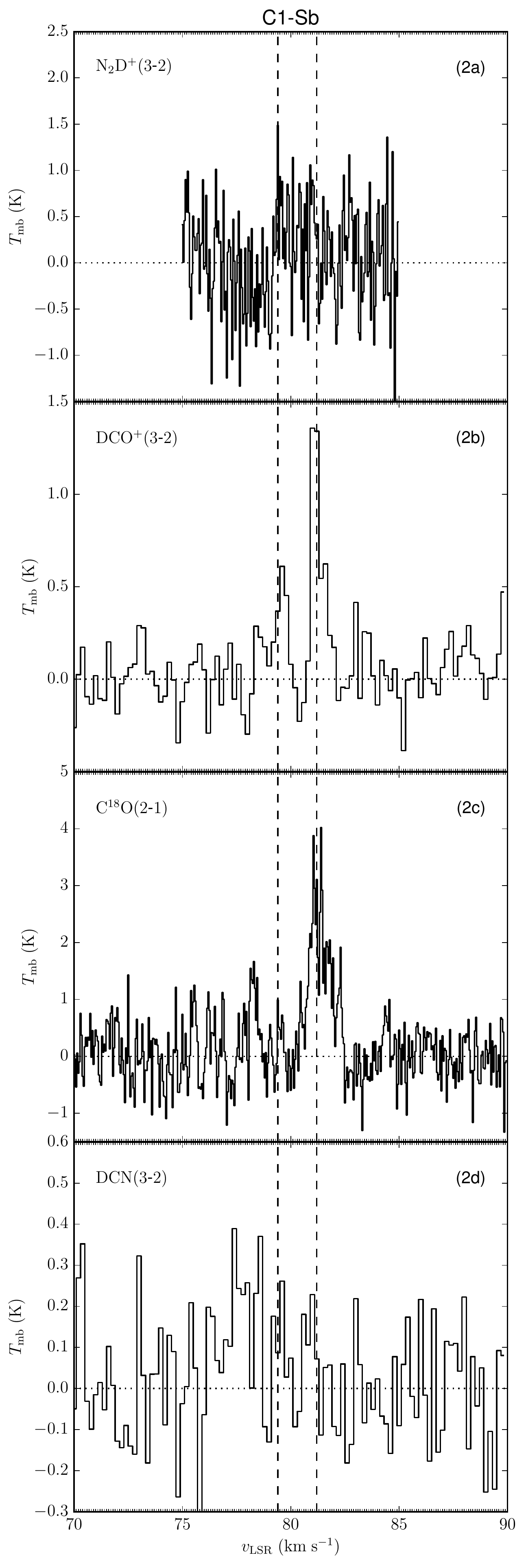}
\plotone{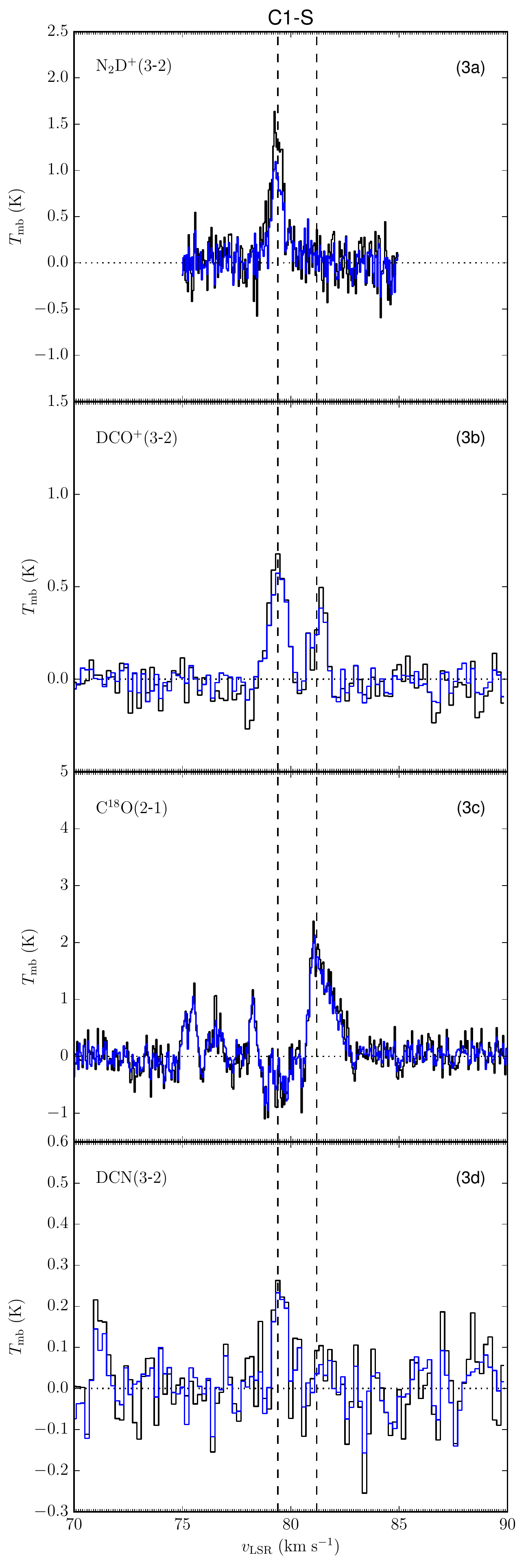}
\caption{
Spectra extracted from the cores (circular regions in
Figure \ref{fig:ntdp1}).  The average main beam temperatures \tmb are
estimated from the mean pixel values (in unit of \jybns) within the
regions defining the cores.  The data cubes are primary beam
corrected.  Columns from left to right: spectra for C1-Sa, C1-Sb and
C1-S.  
The C1-S column shows both C1-S inner (black) and C1-S outer (blue).
Rows from top to bottom: spectra of \ntdpns(3-2),
\dcopns(3-2), \ceions(2-1) and DCN(3-2). The \ntdpns(3-2) and
\ceions(2-1) spectra have a velocity resolution of 0.05 \kmsns. The
\dcopns(3-2) and DCN(3-2) spectra have a velocity resolution of 0.2
\kmsns.  The two vertical dashed lines label the two velocity
components reported in this vicinity by T13. The line at 79.4 \kms
corresponds to the $v_{\rm LSR}$ of the C1-S core in T13 ($v_{\rm
  low}$), while the line at 81.2 \kms corresponds to the C1-N core in
T13 ($v_{\rm high}$). The spectra of C1-Sb are relatively noisy
because of its small size.
\label{fig:spec1}}
\end{figure*}

Figure \ref{fig:spec1} shows the spectra of \ntdpns(3-2),
\dcopns(3-2), \ceions(2-1) and DCN(3-2) from the C1-Sa, C1-Sb and C1-S
inner cores. 
The spectra are averaged over the circular core apertures and
converted to $T_{\rm mb}$. Considering the spectra of the C1-Sa
protostar, one can see that most of its emission from dense gas
tracers is close to the 79.4~\kms velocity component identified in
\ntdpns(3-2) by T13 (hereafter $v_{\rm low}$).  There is a relatively
weak velocity component in \ceions(2-1) at 81.2 \kms (hereafter
$v_{\rm high}$). Like T16, we argue the protostellar core C1-Sa is at
the $v_{\rm low}$ velocity. We note that multiple velocity components
are commonly seen in IRDCs \citep[e.g.,][]{2013MNRAS.428.3425H}.

For the C1-Sb protostar, from the spectra of \dcopns(3-2) and
\ceions(2-1) we see that the main velocity component is at 81.2 \kms
($v_{\rm high}$).  There is a relatively weak component in
\dcopns(2-1) at $v_{\rm low}$ and a hint of a feature in \ntdpns(3-2)
at $v_{\rm low}$.  We argue the protostellar core C1-Sb is at $v_{\rm
  high}$. T16 also suggested a higher radial velocity for C1-Sb based
on the low velocity resolution channel maps of $^{12}$CO(2-1).

The C1-S core, defined by \ntdpns(3-2), is at 79.4 \kms ($v_{\rm
  low}$).  This C1-S component is also seen in \dcopns(3-2) and
DCN(3-2), but not in \ceions(2-1). We also see the $v_{\rm high}$
component in \dcopns(3-2) and \ceions(2-1).

Overall, these spectra suggest that C1-Sa is likely to be part of 
the same cloud enclosing C1-S, while C1-Sb is a core that is part of a
separate larger scale structure, potentially connected to C1-N (T13).
We note that in the C1 region, we generally see these two velocity
components in a variety of datasets. The velocity difference is
typically 1.8 \kmsns.  It remains to be seen whether these components
are interacting, i.e., via a collision, or are simply overlapping in
the plane of the sky but are physically separated in the third spatial
dimension.

\begin{figure*}[htb!]
\epsscale{1.2}
\plotone{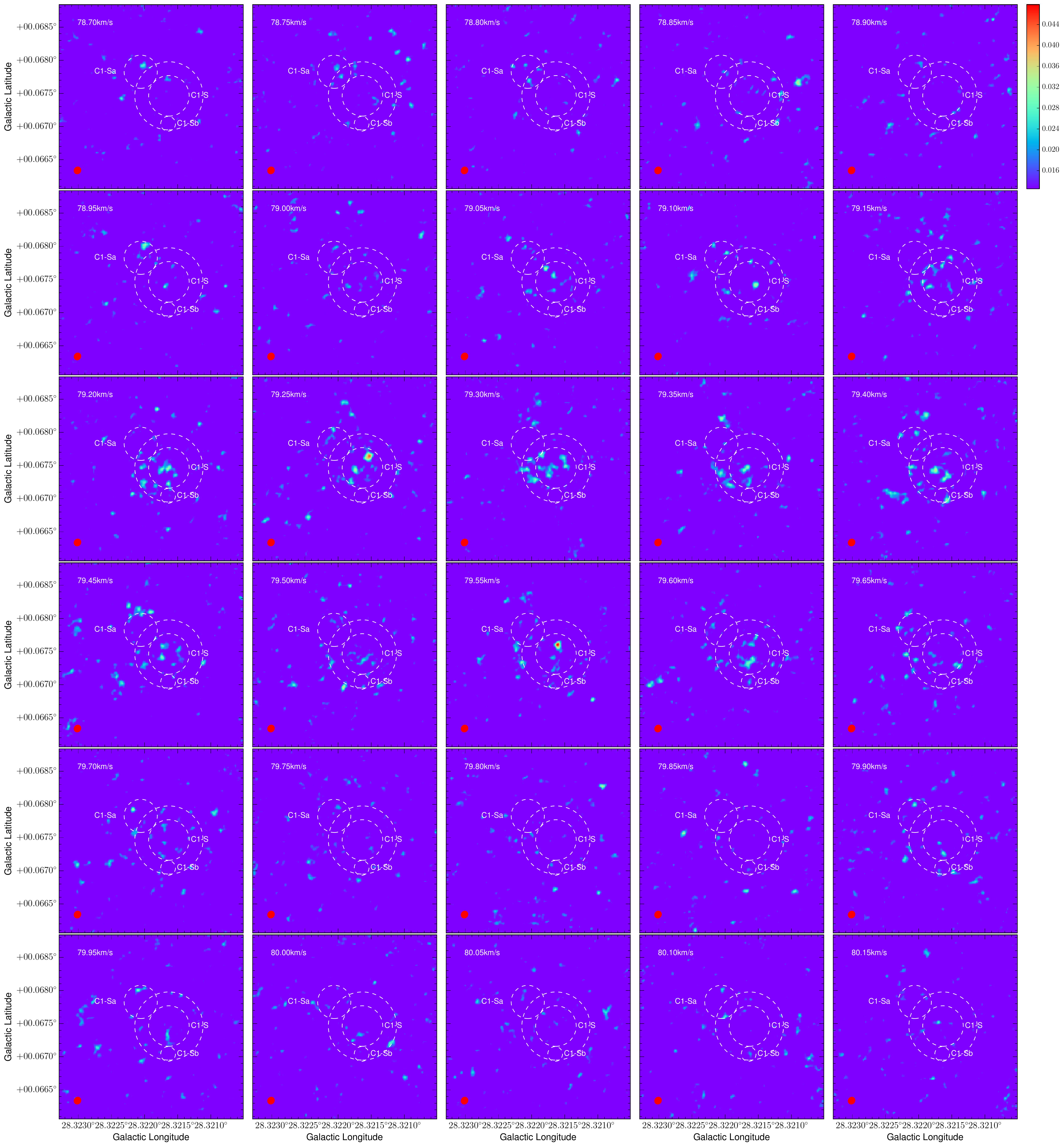}
\caption{
Channel maps of the \ntdpns(3-2) highest resolution data of the C1
region (the same as shown in Figure~\ref{fig:cont1}; core apertures are
marked with dashed circles). The channel velocity is shown in the
upper-left corner of each panel.  The color scheme is from 2$\sigma$
(in this case $\sigma$ = 0.011 \jybns) to the maximum with a linear
scale.  The color bar has a unit of \jybns. The synthesized beam is
shown as a red filled ellipse at lower-left.
\label{fig:chan1}}
\end{figure*}

\begin{figure*}[htb!]
\epsscale{1.2}
\plotone{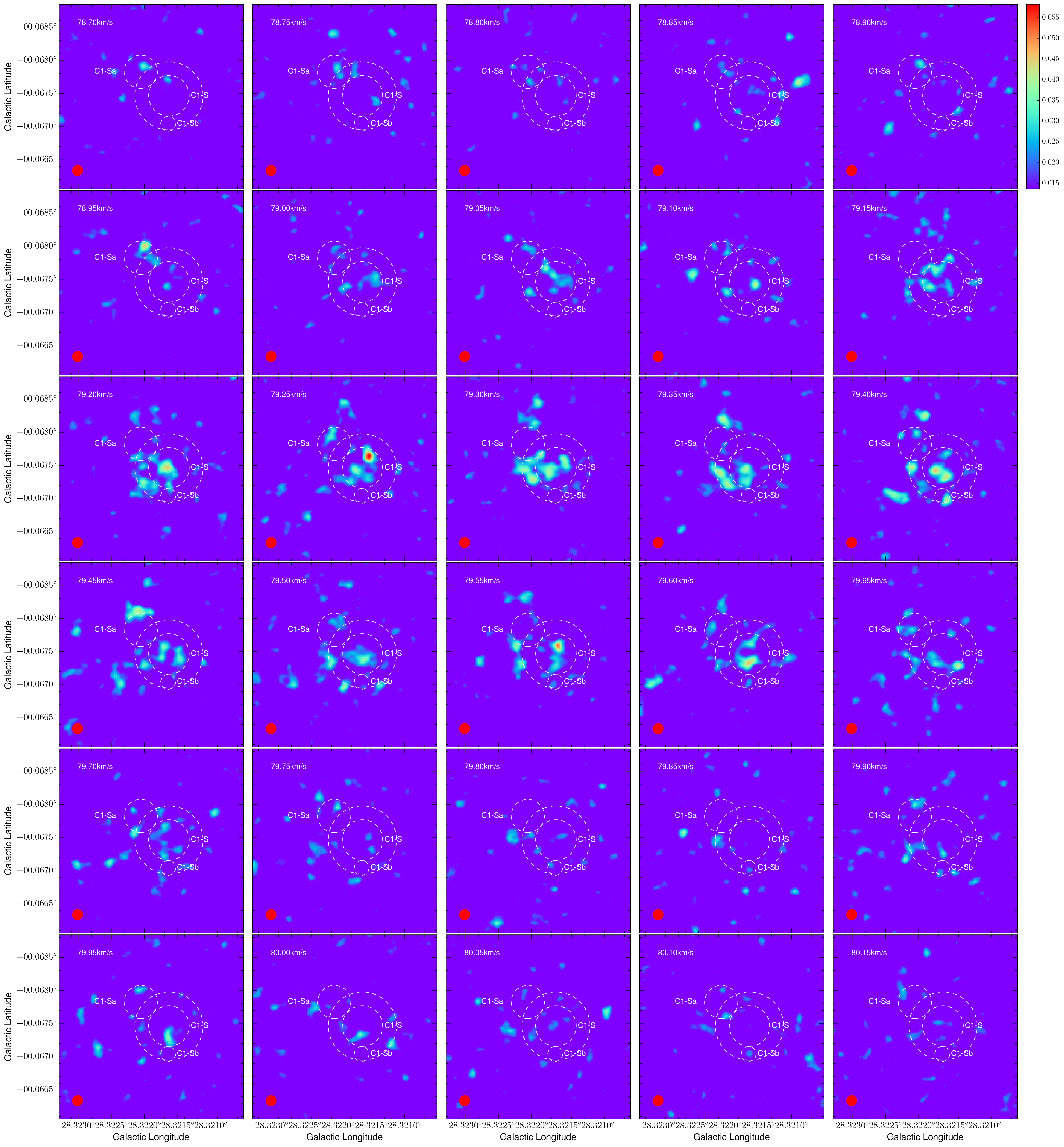}
\caption{
The same as Figure \ref{fig:chan1}, but with the \ntdpns(3-2) data
 uv-tapered to $\sim$ 0.5\arcsec resolution. In this case, $\sigma$ = 0.008
\jybns.
\label{fig:chansmooth1}}
\end{figure*}

Figures~\ref{fig:chan1} and \ref{fig:chansmooth1} show the channel maps
of the \ntdpns(3-2) data, with the latter 
 uv-tapered to $\sim$ 0.5\arcsec
resolution. The three cores C1-Sa, C1-Sb and C1-S are shown with
white dashed circles.  
The velocity range of the channels covers essentially all
of the detected \ntdpns(3-2) emission (above 3$\sigma$). 
\ntdpns(3-2) in C1-Sa is more
prominent in channels $\la$ 79.1 \kmsns, while in C1-S it is more
dominant in channels $\ga$ 79.1 \kmsns.
From 79.1 \kms to 80.0 \kmsns, C1-S is filled with widespread \ntdp
features, with little indication of any large scale velocity gradients.

\begin{figure*}[htb!]
\epsscale{1.2}
\plotone{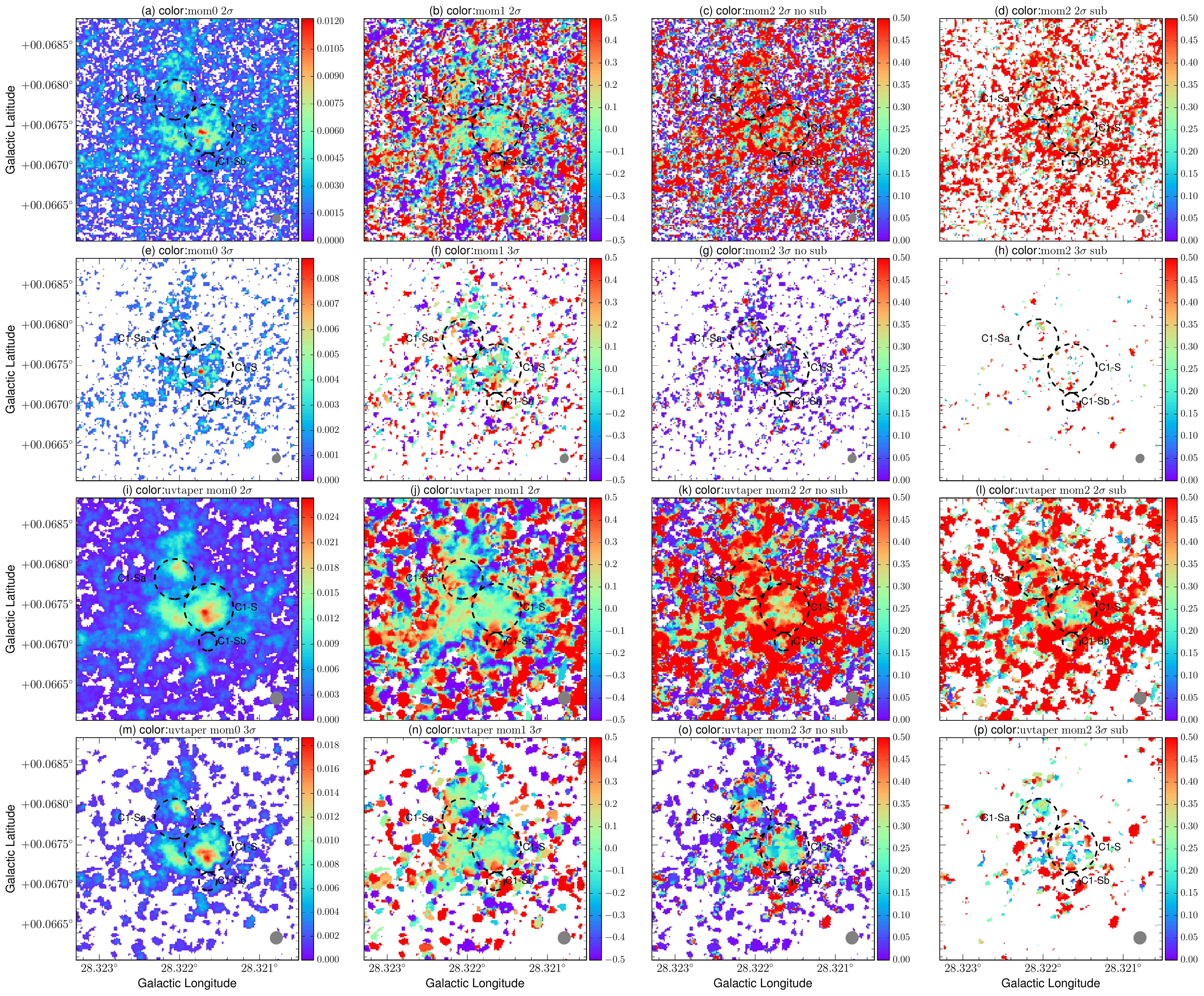}
\caption{
\ntdpns(3-2) moment maps. Columns from left to right: 0th-moment,
1st-moment (relative to 79.4 \kms), 
2nd-moment without intrinsic dispersion subtraction, and
2nd-moment with the \ntdpns(3-2) hyperfine structure dispersion
subtracted.  Rows from top to bottom: 
 combined data with 2$\sigma$
cut, combined data with 3$\sigma$ cut, uv-tapered data (0.5 \arcsec)
with 2$\sigma$ cut, and uv-tapered data with 3$\sigma$ cut. 
All moment
calculations are carried out within the velocity range 79.4$\pm$1.5
\kms to avoid any contribution from the 81.2 \kms component. The
subtraction of the intrinsic dispersion of 0.242 \kms from the
\ntdpns(3-2) hyperfine structure in the 4th column limits the number
of available pixels in the final image.  The beam is shown as the
gray filled ellipse on the lower-right corner.  The color bars for
the 1st column have the unit \jyb \kmsns, for the rest columns \kmsns.
\label{fig:mom012}}
\end{figure*}

To examine this further, Figure \ref{fig:mom012} shows the 1st-moment
map of the combined \ntdpns(3-2) data, with velocity measurements
shown for those pixels with a 2$\sigma$ and 3$\sigma$ detection
of the species, i.e., to illustrate the effect of the choice of
  this threshold. C1-S shows a quite coherent structure around 79.4
\kmsns.  No clear sign of global rotation is seen in C1-S.  There is a
stronger velocity gradient seen across C1-Sa, equivalent to about a
0.8~\kmsns change across the diameter of the core
 (see Figure \ref{fig:mom012}(j) and (n)). 
The direction of the gradient (E-W) is consistent with the continuum
elongation.  
Figure \ref{fig:mom012} shows the 2nd-moment map of the combined \ntdp 
data (the last two columns). 
 In the fourth column, we subtract the intrinsic dispersion of the 
\ntdpns(3-2) hyperfine structure in quadrature from the 2nd-moment map
(calculated at 2$\sigma$ and 3$\sigma$).
Consequently, the number of available pixels is
limited by the sensitivity. 
C1-Sa shows moderately enhanced
2nd-moment, which may be caused by its outflow (\S\ref{subsec:co}).

\begin{figure}[htb!]
\epsscale{1}
\plotone{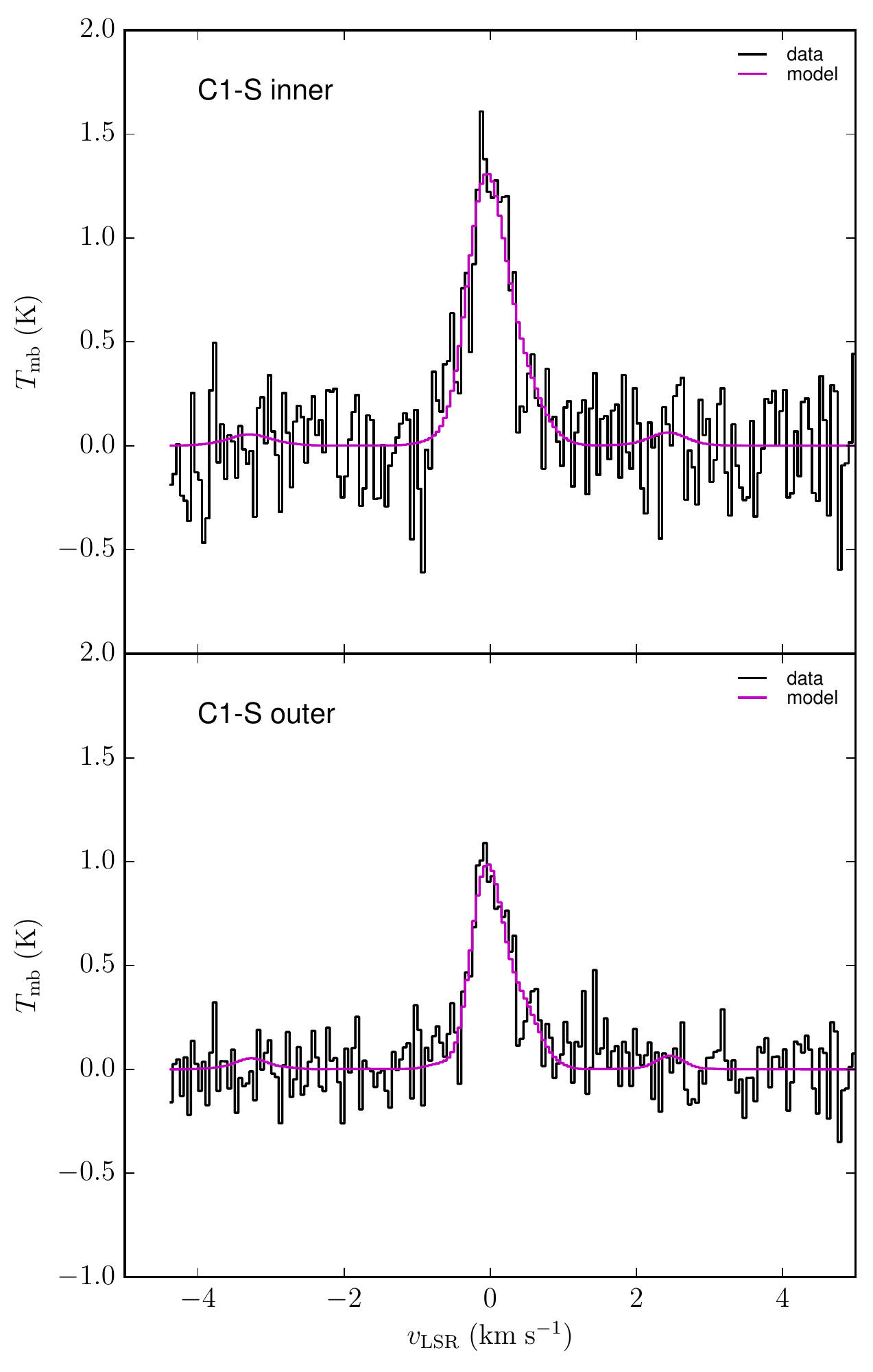}
\caption{
\ntdpns(3-2) core spectra (black) and their HFS fit results (magenta).
All spectra are in the rest frame of 79.4 \kmsns. The spectral
resolution is 0.05 \kmsns.
\label{fig:corefit}}
\end{figure}

We performed hyperfine-structure spectral fitting to the \ntdpns(3-2)
spectrum using the HFS method of the CLASS
software package\footnote{http://www.iram.fr/IRAMFR/GILDAS} 
 in Figure \ref{fig:corefit}, assuming the line is optically thin
  \citep[see the optical depth check in][]{2016ApJ...821...94K}.  The
thermal component of \ntdp is subtracted in quadrature from the line
dispersion, and the thermal component of H$_2$ gas (assuming
$10\pm3$~K) is added back in quadrature to obtain the total velocity
dispersion of the core as derived from the \ntdp observations
($\sigma_{\rm N_2D^+}$) (see T13).  These results are listed in Table
\ref{tab:2}. The observed line profiles of \ntdp are dominated by
turbulent motions, although the total velocity dispersions are under
$0.4\:{\rm km\:s^{-1}}$.

\begin{figure}[htb!]
\epsscale{1.}
\plotone{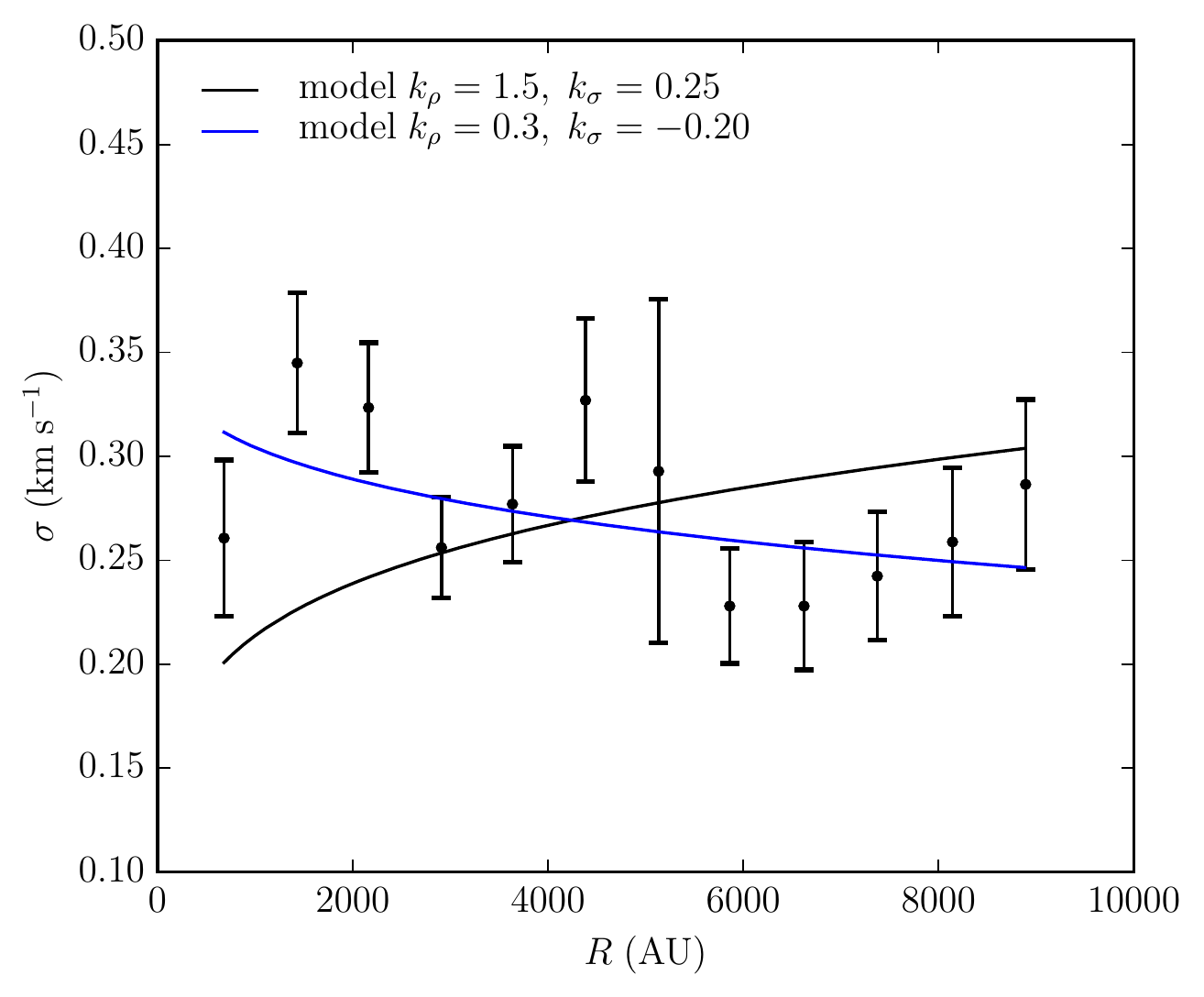}
\caption{
 Velocity dispersion $\sigma$ as a function of projected
radius, $R$, for C1-S outer. The observational measurement is averaged within
concentric annuli at a step of beam size.  Within each annulus, the observed
$\sigma_{\rm N_2D^+}$ is derived following \S\ref{subsubsection:ck}.
The black model curve shows the 2D
projection (mass-weighted sum in quadrature) of 
$\sigma(r)~\propto~r^{k_\sigma}$ (${k_\sigma}=0.25$)
with the density profile $\rho(r)~\propto~r^{-k_\rho}$ ($k_\rho=1.5$), 
following the fiducial MT03 model. The profile is normalized based on a $\chi^2$
minimization to the observation so that the $\sigma$ at core surface
(9250 AU) is $\sigma_s=0.306~\rm km~s^{-1}$.  The blue model curve
shows the model with a shallower density profile
$\rho(r)~\propto~r^{-0.3}$ and the best fit $k_\sigma=-0.201$.  In this
case, $\sigma_s=0.245~\rm km~s^{-1}$.
\label{fig:radial}}
\end{figure}

Figure \ref{fig:radial} shows the C1-S outer core velocity dispersion
$\sigma$ as a function of projected radius, $R$.  
The core is dissected into the same set of annuli as in Figure
\ref{fig:massradial}.  Within each annulus, we extract the averaged
\ntdpns(3-2) flux density and form a spectrum. Then we fit the
hyperfine structure of \ntdpns(3-2) 
 and derive $\sigma_{\rm N_2D^+}$, as described above.

 We first model the core velocity dispersion following the MT03
  model, where the core volume density follows
  $\rho(r)~\propto~r^{-k_\rho}$ and the velocity dispersion follows
  $\sigma(r)~\propto~r^{k_\sigma}$.  The fiducial MT03 model has
  $k_\rho=1.5$ and $k_\sigma=0.25$.  We project the $\sigma(r)$ to the
  sky plane by summing the line-of-sight mass-weighted $\sigma(r)$ in
  quadrature (this is the same as the emission-weighted averaging in
  the optically thin, isothermal limit). Finally we normalize the 2D
  $\sigma-R$ model based on a $\chi^2$ minimization to the data
  (converted to $\sigma_{\rm N_2D^+}$, following the methods described
  above).  The model curve (black) is shown in Figure
  \ref{fig:radial}.  It has a velocity dispersion at core surface of
  $\sigma_s=0.306~\rm km~s^{-1}$. We also try a shallower density
  profile with $k_\rho=0.3$.  This time we allow $k_\sigma$ to vary,
  finding a best-fit result $k_\sigma=-0.201$ and $\sigma_s=0.245~\rm
  km~s^{-1}$.

 These results indicate that velocity dispersion is fairly constant
with radius in the C1-S core. The fiducial turbulent core model is
consistent with the observed velocity dispersion versus projected
radius profile, although its continuum flux profile is too
concentrated (unless one invokes strong temperature and/or opacity
gradients). A flatter density profile, i.e., $k_\rho=0.3$, which is
more favored by the continuum flux profile, still implies a relatively
flat velocity dispersion profile, although now declining with
radius. However, these results assume a constant abundance of \ntdp in
the core, which is a caveat that needs to be borne in mind.

\subsubsection{Core Dynamics}\label{subsubsec:dynamics}

We follow T13 to measure the C1-S core properties at both the inner
and outer scales and investigate its dynamics.  We first use the MIREX
image from \citet{2014ApJ...782L..30B} to estimate the mass surface
density of the core ($\Sigma_c$) and its clump envelope ($\Sigma_{\rm
  cl,MIREX}$), evaluated from an annulus extending out to twice the
core radius. However, we note that the C1-S region may be
``saturated'' in the MIREX map, i.e., it may only provide a lower
limit to the actual mass surface density \citep[see discussion
  in][]{2012ApJ...754....5B,2014ApJ...782L..30B,2014ApJ...780L..29L}.
For the C1-S core we find $\Sigma_{\rm c,MIREX}\simeq0.5\:{\rm
  g\:cm^{-2}}$ for both the inner and outer core regions. We also find
quite similar values for the clump mass surface density via this
method, i.e., $\Sigma_{\rm cl,MIREX}$. This indicates that saturation
in the map is indeed a problem. Another potential difficulty with
estimates based on the MIREX map is that its angular resolution is
only 2\arcsec.

As a result of these potential problems with the MIREX map of this
region, we also measure mass surface densities of the core and clump
envelopes based on their mm continuum emission, i.e., 
$\Sigma_{\rm c,mm,tot}$ and $\Sigma_{\rm cl,mm}$. 
The subscript ``tot" means no
envelope subtraction (see below). Following T13, we adopt $T =
10\pm3\:$K for the core, which results in lower and upper limits
denoted as sub- and superscripts. For the clump we adopt
$T=15\pm5\:$K, i.e., with such temperatures being expected to be
relevant for larger-scale regions in IRDCs \citep[e.g.,][see also 
Fig.~\ref{fig:tkin}]{2006A&A...450..569P}.
With these methods we find that C1-S core
has $\Sigma_{\rm c,mm,tot}\simeq 2.41$ and $1.96\:{\rm g\:cm^{-2}}$ at its
inner and outer scales, respectively. The surrounding clump envelopes
around these apertures have $\Sigma_{\rm cl,mm}\simeq 0.80$ and
$0.34\:{\rm g\:cm^{-2}}$, respectively.  Given these results, we focus
our attention on core properties for the dynamical analysis using the
mm continuum derived mass surface densities.

We consider two cases for estimating core physical properties: (1) no
envelope subtraction, i.e., using total mm continuum fluxes, for which
we use the subscript ``tot'', e.g., $S_{\rm c,mm,tot}$, 
$\Sigma_{\rm c,mm,tot}$, $M_{\rm
  c,mm, tot}$, etc.; (2) with envelope subtraction, for which we use
variable names with no additional subscript, e.g., $\Sigma_{\rm
  c,mm}$, $M_{\rm c,mm}$, etc. The resulting physical properties of
C1-S, including volume densities (assuming spherical geometry and
  adopting a mean molecular weight of $2.33 m_{\rm H}$), are listed
in Table \ref{tab:2}.  With the fiducial dust temperature estimates,
the derived masses of C1-S inner are $M_{\rm c,mm,tot}=25.5\:M_\odot$
and $M_{\rm c,mm}=17.1\:M_\odot$, i.e., background subtraction makes a
substantial difference in the mass estimate. For C1-S outer, the
masses are $M_{\rm c,mm,tot}=58.8\:M_\odot$ and $M_{\rm
  c,mm}=48.8\:M_\odot$, i.e., the effects of background subtraction
are more modest (in a relative sense). The implied volume
densities in C1-S, also listed in Table~\ref{tab:2}, are $n_{\rm H}
\sim$ several $\times10^6\:{\rm cm^{-3}}$.

\begin{table}
\centering
\begin{threeparttable}
\caption{Physical properties of C1-S}\label{tab:2}
\begin{tabular}{ccc}
\hline {{Core property (\% error)}} & {{C1-S inner}} & {{C1-S outer}}\\
\hline
Angular radius, $\theta_c$ (\arcsec) & 1.10 & 1.85 \\  
$d$ (kpc) (20\%) & 5.0 & 5.0 \\  
$R_{\rm c}$ (0.01~pc) (20\%) & 2.67 & 4.48 \\  
\hline
\hline
$V_{\rm LSR,N_2D^+}$ ${\rm (km\:s^{-1})}$ & 79.30$\pm$0.0200 & 79.30$\pm$0.0200 \\  
$\sigma_{\rm N_2D^+,obs}$ ${\rm (km\:s^{-1})}$ & 0.242$\pm$0.0251 & 0.212$\pm$0.0179 \\  
$\sigma_{\rm N_2D^+,nt}$ ${\rm (km\:s^{-1})}$  & 0.236$\pm$0.0257 & 0.205$\pm$0.0185 \\  
$\sigma_{\rm N_2D^+}$ ${\rm (km\:s^{-1})}$     & 0.302$\pm$0.0267 & 0.279$\pm$0.0231 \\  
\hline
\hline
$\Sigma_{\rm cl,MIREX}$ ${\rm (g\:cm^{-2})}$ (30\%) & 0.490 & 0.445 \\  
\hline
$S_{\rm cl,mm}$ (mJy) & 17.5$\pm$0.577 & 23.8$\pm$1.04 \\  
$S_{\rm cl,mm}/\Omega$ (MJy/sr) & 78.7$\pm$2.60 & 33.1$\pm$1.44 \\  
$\Sigma_{\rm cl,mm}$ ${\rm (g\:cm^{-2})}$ & 0.798$_{0.541}^{1.48}$ & 0.336$_{0.228}^{0.624}$ \\  
\hline
\hline
$\Sigma_{\rm c,MIREX}$ ${\rm (g\:cm^{-2})}$ (30\%) & 0.492 & 0.491 \\  
\hline
$M_{\rm c,MIREX}$ $(M_\odot)$ (50\%) & 5.22 & 14.7 \\  
\hline
$n_{\rm H,c,MIREX}$ $(10^5{\rm cm}^{-3})$ (36\%) & 19.0 & 11.3 \\  
\hline
\hline
$S_{\rm c,mm,tot}$ (mJy) & 11.4$\pm$0.365 & 23.2$\pm$0.577 \\  
$S_{\rm c,mm,tot}/\Omega$ (MJy/sr) & 128$\pm$4.10 & 104$\pm$2.60 \\  
\hline
$\Sigma_{\rm c,mm,tot}$ ${\rm (g\:cm^{-2})}$ & 2.41$_{1.43}^{4.83}$ & 1.96$_{1.16}^{3.93}$ \\  
\hline
$M_{\rm c,mm,tot}$ $(M_\odot)$ & 25.5$_{11.6}^{53.5}$ & 58.8$_{26.8}^{123}$ \\  
\hline
$n_{\rm H,c,mm,tot}$ $(10^5{\rm cm}^{-3})$ & 93.0$_{51.6}^{190}$ & 45.1$_{25.0}^{92.1}$ \\  
\hline
\hline
$\Sigma_{\rm c,mm}$ ${\rm (g\:cm^{-2})}$ & 1.61$_{0.628}^{4.03}$ & 1.63$_{0.826}^{3.60}$ \\  
\hline
$M_{\rm c,mm}$ $(M_\odot)$ & 17.1$_{4.97}^{43.7}$ & 48.8$_{19.0}^{112}$ \\  
\hline
$n_{\rm H,c,mm}$ $(10^6{\rm cm}^{-3})$ & 6.22$_{2.24}^{15.8}$ & 3.73$_{1.77}^{8.39}$ \\  
\hline
\hline
\end{tabular}
\begin{tablenotes}
\small
\item The superscripts (low temperature) and subscripts (high
  temperature) indicate the variations resulting from dust temperature
  uncertainties (7 to 13~K for core; 10 to 20~K for clump). 
\end{tablenotes}
\end{threeparttable}
\end{table}

Now with estimates of the radii, masses and velocity dispersions of
the C1-S inner and outer core structures, along with their clump
envelope mass surface densities, we are able to compare these to
predictions of the Turbulent Core Model of MT03. Our general
methodological approach follows that of T13. The mass-weighted
  average  velocity dispersion of a virialized core, including
pressure equilibrium with its surroundings, is given, in the fiducial
case by (MT03; T13):
\begin{equation}
\sigma_{\rm c,vir} \rightarrow 1.09
\left(\frac{\phi_B}{2.8}\right)^{-3/8}\left(\frac{M_c}{60
  M_\odot}\right)^{1/4} \left(\frac{\Sigma_{\rm cl}}{1\:{\rm
    g\:cm^{-2}}}\right)^{1/4}\:{\rm km\:s^{-1}},
\label{eq:vir}
\end{equation}
where $\phi_B=1.3+1.5 m_A^{-2}\rightarrow 2.8$ is a dimensionless
parameter that accounts for the effects of magnetic fields,
$m_A\equiv\sqrt{3}\sigma/v_A\rightarrow 1$ is the Alfv\'en Mach
number and $v_A=B/\sqrt{4\pi\rho}$ is the Alfv\'en speed. The fiducial
case considered by MT03 (indicated by the $\rightarrow$ values, above)
involved approximately equal support in the core by turbulence and
large-scale magnetic fields. If magnetic fields play a more important
role, then this is represented by a smaller Alfv\'en Mach number,
i.e., sub-Alfv\'enic turbulence, and a larger value of $\phi_B$. In
this case, a virial equilibrium core would have a smaller turbulent
line width.

Thus our basic procedure is to evaluate $\sigma_{\rm c,vir}$ using
equation (\ref{eq:vir}) and compare it with the observed 1D velocity
dispersion inferred from the \ntdpns(3-2) line, $\sigma_{\rm
  N_2D^+}$. We do this for the C1-S inner and outer scales and for the
case of no clump envelope background subtraction and with such
background subtraction. These results are listed in Table~\ref{tab:3}.
The ratio $\sigma_{\rm N_2D^+}/\sigma_{\rm c,vir,mm}$ is found to be
approximately equal to 0.4 for both C1-S inner and outer, which is
similar to the results of T13 based on the lower resolution
observations of C1-S and of \citet{2017ApJ...834..193K} 
for other \ntdp cores in IRDCs. 

There are two possible interpretations for these results. The first
possibility is that the core really is in a sub-virial state, i.e., it
is on the verge of undergoing fast collapse because it lacks
sufficient internal pressure support. Evaluation of the \citet{1992ApJ...395..140B} 
virial parameter (see Table~\ref{tab:3}), which is often
used to assess the dynamical state of cores, would also seem to
indicate such a situation. However, the high densities of the core
imply short free-fall times (also listed in Table~\ref{tab:3}), which
are $\sim2\times10^4\:$yr. Gas in the core would be expected to
relatively quickly acquire infall velocities, which would approach the
free-fall speed and thus give the core an apparent velocity dispersion
that is comparable to that of virial equilbrium. It seems unlikely
that all the core material is slowly moving and on the verge of fast
collapse, especially on the different scales of C1-S inner and
outer. In addition, the astrochemical modeling of the deuteration
process that increases the abundance of \ntdp compared to \nthp is
thought to take a relatively long time compared to the local free-fall
time. The results of \citet{2016ApJ...821...94K,2016ApJ...833..274G}
indicate that the C1-S core should be contracting at a rate smaller
than 1/3 of that of free-fall collapse in order to have enough time to
reach the observed levels of deuteration.

The second possibility is that the C1-S core is quite close to a state
of virial equilibrium and is only undergoing relatively slow
contraction compared to that of free-fall collapse. This would then
require stronger magnetic fields compared to those of the fiducial
($m_A=1$) turbulent core model. The values of these fiducial $B$-field
strengths are about 0.5~mG on the scale of C1-S outer and 0.8~mG at
C1-S inner. The values of $\phi_B$ to achieve virial equilibrium,
$\phi_{\rm B,vir}$, are several times larger than that of the fiducial
case (Table~\ref{tab:3}). These correspond to conditions of
sub-Alfv\'enic turbulence with $m_A\simeq0.2$ and magnetic field
strengths of $\sim 2$ to 3~mG. We note that such $B$-field strengths
are consistent with the values predicted by the empirical relation for
median values of \citet{2010ApJ...725..466C}, $B_{\rm med}\simeq0.22
(n_{\rm H}/10^5\:{\rm cm^{-3}})^{0.65}$ (valid for $n_{\rm
  H}>300\:{\rm cm^{-3}}$), given the observed densities of C1-S inner
and outer. This relation predicts $B$-field strengths of 3.2 and
2.7~mG using the envelope-subtracted densities of C1-S inner and
outer, respectively. Finally, such values of $B$-field strengths can
also help to explain why C1-S does not appear to have fragmented
significantly, i.e., based on its mm continuum
morphology\footnote{Note the order unity fluctuations in $\rm N_2D^+$
  column density within C1-S are easily within the range expected from
  abundance variations (Goodson et al. 2016). Furthermore, we have
  carried out a similar dynamical analysis on local $\rm N_2D^+$ peaks
  and find these to be relatively more gravitationally stable compared
  to C1-S at its inner and outer scales.}. Table~3 lists the magnetic
field strengths, $B_{\rm c,crit}$, that would be needed for the
magnetic critical mass \citet{1992ApJ...395..140B} to equal the
observed core masses: at the scale of C1-S outer we see that $B_{\rm
  c,crit}\simeq B_{\rm c,vir}\simeq 2\:$mG.

In summary, given the above results, we conclude that the second case
of core dynamics regulated by relatively strong, $\sim 2\:$mG,
magnetic fields appears to be the more likely scenario. Such magnetic
field strengths are reasonable given the observed densities. They help
explain the fragmentation scale of C1-S outer, i.e., $\sim
50\:M_\odot$. They would allow C1-S outer to be virialized and thus
potentially relatively old compared to its free-fall time, which helps
to explain its observed high level of deuteration of \nthp, i.e., high
abundance of \ntdpns. We return to this point in \S\ref{S:discussion},
where we consider the implications of the observed C1-S properties for
astrochemical models. The predictions of there being dynamically
important magnetic fields are: (1) strong Zeeman splitting of species,
such as CN, if they are present in the gas phase within the core; (2)
ordered dust continuum emission polarization angles, assuming dust
grains can align with the $B$-fields; (3) relatively small infall
rates compared to free-fall. Again, we return to discussion of these
predictions below in \S\ref{S:discussion}.

\begin{table}
\centering
\begin{threeparttable}
\caption{Dynamical properties of C1-S}\label{tab:3}
\begin{tabular}{ccc}
\hline {{Core property}} & {{C1-S inner}} & {{C1-S outer}} \\
\hline
$R_{\rm c}$ (0.01~pc) & 2.67 & 4.48 \\  
$\sigma_{\rm N_2D^+}$ ${\rm (km\:s^{-1})}$ & 0.302$\pm$0.0267 & 0.279$\pm$0.0231 \\  
$\Sigma_{\rm cl,mm}$ ${\rm (g\:cm^{-2})}$ & 0.798$_{0.541}^{1.48}$ & 0.336$_{0.228}^{0.624}$ \\  
\hline
\hline
$M_{\rm c,mm,tot}$ $(M_\odot)$ & 25.5$_{11.6}^{53.5}$ & 58.8$_{26.8}^{123}$ \\  
$M_{\rm c,mm}$ $(M_\odot)$ & 17.1$_{4.97}^{43.7}$ & 48.8$_{19.0}^{112}$ \\  
\hline
$\sigma_{\rm c,vir,mm,tot}$ ${\rm (km\:s^{-1})}$ & 0.832$_{0.606}^{1.03}$ & 0.826$_{0.602}^{1.02}$ \\  
$\sigma_{\rm c,vir,mm}$ ${\rm (km\:s^{-1})}$ & 0.752$_{0.500}^{0.973}$ & 0.788$_{0.557}^{0.994}$ \\  
\hline
$\sigma_{\rm N_2D^+}/\sigma_{\rm c,vir,mm,tot}$  & 0.363$_{0.293}^{0.499}$ & 0.337$_{0.272}^{0.463}$ \\  
$\sigma_{\rm N_2D^+}/\sigma_{\rm c,vir,mm}$  & 0.402$_{0.311}^{0.604}$ & 0.354$_{0.280}^{0.500}$ \\  
\hline
$n_{\rm H,c,mm,tot}$ $(10^5{\rm cm}^{-3})$ & 93.0$_{51.6}^{190}$ & 45.1$_{25.0}^{92.1}$ \\  
$n_{\rm H,c,mm}$ $(10^5{\rm cm}^{-3})$ & 62.2$_{22.4}^{158}$ & 37.3$_{17.7}^{83.9}$ \\  
\hline
$t_{\rm c,ff,tot}$ $(10^5{\rm yr})$\tablenotemark{a} & 0.143$_{0.100}^{0.192}$ & 0.206$_{0.144}^{0.276}$ \\  
$t_{\rm c,ff}$ $(10^5{\rm yr})$\tablenotemark{a} & 0.175$_{0.110}^{0.291}$ & 0.226$_{0.151}^{0.328}$ \\  
\hline
$\alpha_{\rm c,tot}\equiv 5\sigma_{\rm N_2D^+}^2 R_{c}/(G M_{\rm c,mm,tot})$\tablenotemark{b}  & 0.110$_{0.0457}^{0.245}$ & 0.0683$_{0.0286}^{0.152}$ \\  
$\alpha_{c}\equiv 5\sigma_{\rm N_2D^+}^2 R_{c}/(G M_{\rm c,mm})$\tablenotemark{b}  & 0.165$_{0.0554}^{0.568}$ & 0.0824$_{0.0315}^{0.213}$ \\
\hline
$B_{c,tot}$ (mG) ($m_A=1$) & 0.865$_{0.632}^{1.24}$ & 0.555$_{0.406}^{0.798}$ \\  
$B_c$ (mG) ($m_A=1$) & 0.708$_{0.418}^{1.13}$ & 0.506$_{0.343}^{0.761}$ \\  
\hline
$\phi_{\rm B,vir,tot}$ & 41.7$_{17.9}^{73.7}$ & 50.8$_{21.8}^{89.8}$ \\  
$\phi_{\rm B,vir}$ & 31.9$_{10.7}^{63.3}$ & 44.8$_{17.8}^{83.2}$ \\  
\hline
$m_{\rm A,vir,tot}$ & 0.193$_{0.144}^{0.301}$ & 0.174$_{0.130}^{0.270}$ \\  
$m_{\rm A,vir}$ & 0.221$_{0.156}^{0.399}$ & 0.186$_{0.135}^{0.302}$ \\  
\hline
$B_{\rm c,vir,tot}$ (mG) & 4.49$_{2.47}^{6.98}$ & 3.19$_{1.77}^{4.95}$ \\  
$B_{\rm c,vir}$ (mG) & 3.19$_{1.27}^{5.54}$ & 2.72$_{1.36}^{4.43}$ \\  
\hline
$B_{\rm c,crit,tot}$ (mG) & 3.24$_{1.95}^{5.42}$ & 2.64$_{1.59}^{4.42}$ \\  
$B_{\rm c,crit}$ (mG) & 2.17$_{0.873}^{4.19}$ & 2.19$_{1.15}^{3.90}$ \\  
\hline
\hline
\end{tabular}
\begin{tablenotes}
\small
\item The superscripts (low temperature) and subscripts (high
  temperature) indicate the variations resulting from dust temperature
  uncertainties (7 to 13~K for core; 10 to 20~K for clump). 
\item $^a$ Core free-fall time, $t_{\rm c,ff} = [3\pi/(32G\rho_c)]^{1/2} = 1.38 \times 10^5 (n_{\rm H,c,mm}/10^5\:{\rm cm^{-3}})^{-1/2}\:{\rm yr}$.
\item $^b$ Virial parameter \citep{1992ApJ...395..140B}.
\end{tablenotes}
\end{threeparttable}
\end{table}

\subsection{CO Outflows}\label{subsec:co}

\begin{figure*}[htb!]
\epsscale{1.}
\plotone{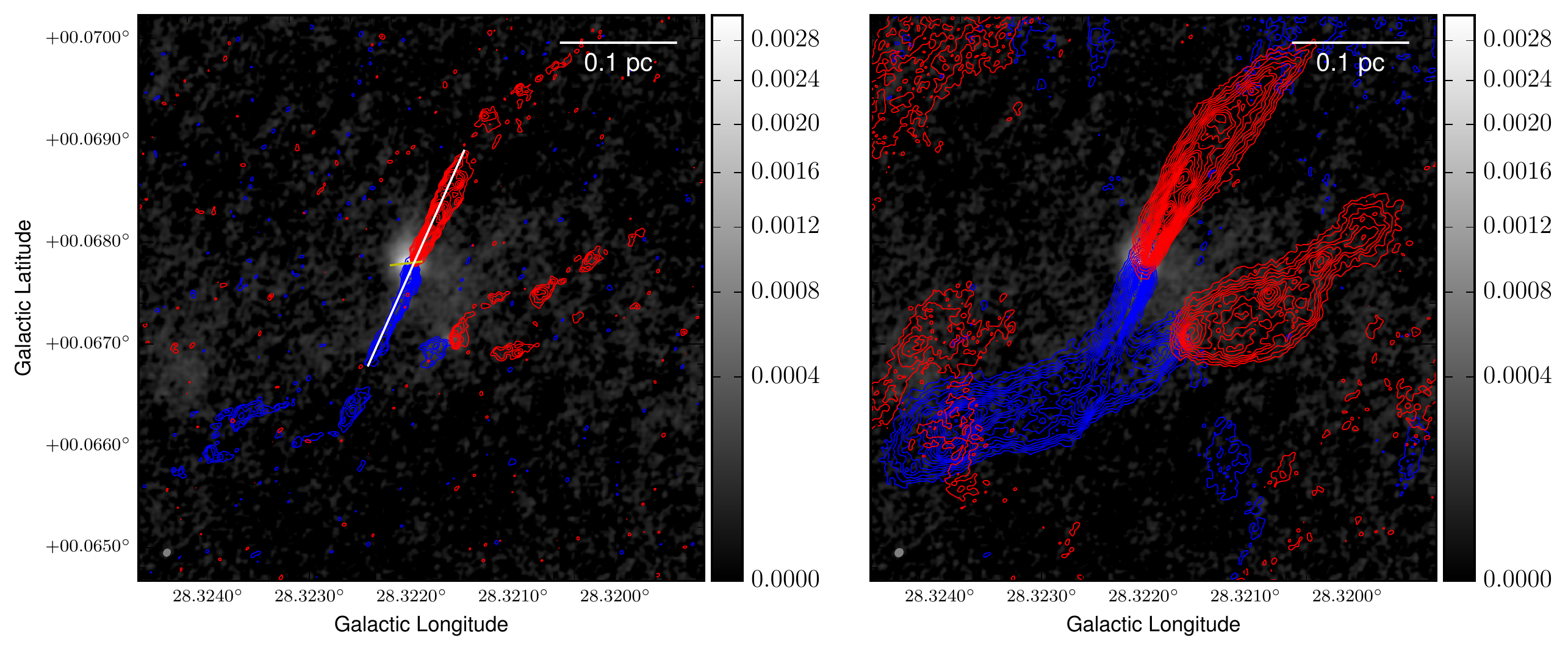}
\caption{
CO(2-1) 0th-moment contours overlaid on the continuum images
(grayscale with unit of \jybns, showing combined data from
Figure \ref{fig:cont1}). Blue contours show emission integrated from
33.8 to 72.8 \kmsns; red contours show emission integrated from 85.8
to 124.8 \kmsns. The displayed contour levels are
5, 7.5, 10, 12.5$\sigma$... 
{\it (a) Left panel:} Outflow contours from extended-configuration
only data (max contour level 25$\sigma$ with $\sigma$ = 16 mJy beam$^{-1}$ \kmsns). 
{\it (b) Right panel:} Outflow contours from combined data 
 (max contour level 45$\sigma$ with with $\sigma$ = 12 mJy beam$^{-1}$ \kmsns).
The synthesized beams are shown in the lower left corners as filled
gray ellipses.
\label{fig:co}}
\end{figure*}

Figure \ref{fig:co} shows the new high-resolution CO(2-1) data, which
can be compared to the compact configuration results presented by T16.
In the figure we show both the extended-only data
(Figure~\ref{fig:co}a), which emphasizes the finest structures, and
the combined data (Figure~\ref{fig:co}b). In panel (a), we can see a
very narrow and collimated bipolar outflow launched from C1-Sa. We
draw a white line to roughly represent the outflow axis. We also draw
a yellow line to show the approximate orientation of the continuum
elongation.  The position angle between the two lines is about
60$^\circ$, which means the angle between the normal direction to the
continuum elongation and the outflow axis is approximately
30$^\circ$. Also relevant is that the origin of the outflow appears to
be slightly offset from the continuum peak (by one beam).
The outflow is only marginally resolved in the direction perpendicular
to the flow axis (by $\sim3$ beams). An upper limit on the half
opening angle of the outflow cavity is estimated to be
$\lesssim10^\circ$.  One can see clumpy structures in both the blue
and red lobes. The redshifted outflow appears to change the
orientation of its flow axis by a small angle, $\sim10^\circ$, once it
is about 0.1~pc away from the protostar. The blueshifted side does not
show evidence for such a deflection, however, its overlap with the
C1-Sb outflow makes this harder to discern. The observed extent of the
outflow appears to be set by sensitivity, rather than being a real,
physical limit.

The outflow from C1-Sb shows a wider opening angle at its base. The
cavity walls, i.e., outflow lobe boundaries, are apparent in the
extended configuration image. The blue lobe overlaps with that from
C1-Sa, but it is not clear if this involves a physical interaction or
is simply a projection effect.

There appear to be some high-velocity CO(2-1) emission features in the
wider region. However, some of these features may be artifacts
resulting from imperfect cleaning of the image, given the strong
sources of the C1-Sa and C1-Sb outflows. Or these features could be
outflows from other nearby protostars or other ambient high-velocity
gas in the IRDC or projected along the line of sight.

Finally and most importantly for the purposes of this paper, neither
panel (a) or panel (b) of Figure~\ref{fig:co} show any hint of CO
outflows from C1-S. This is further evidence that the C1-S core is
likely to be starless.

\subsection{Ancillary Molecular Lines}\label{subsec:anci}

\begin{figure*}[htb!]
\epsscale{1.}
\plotone{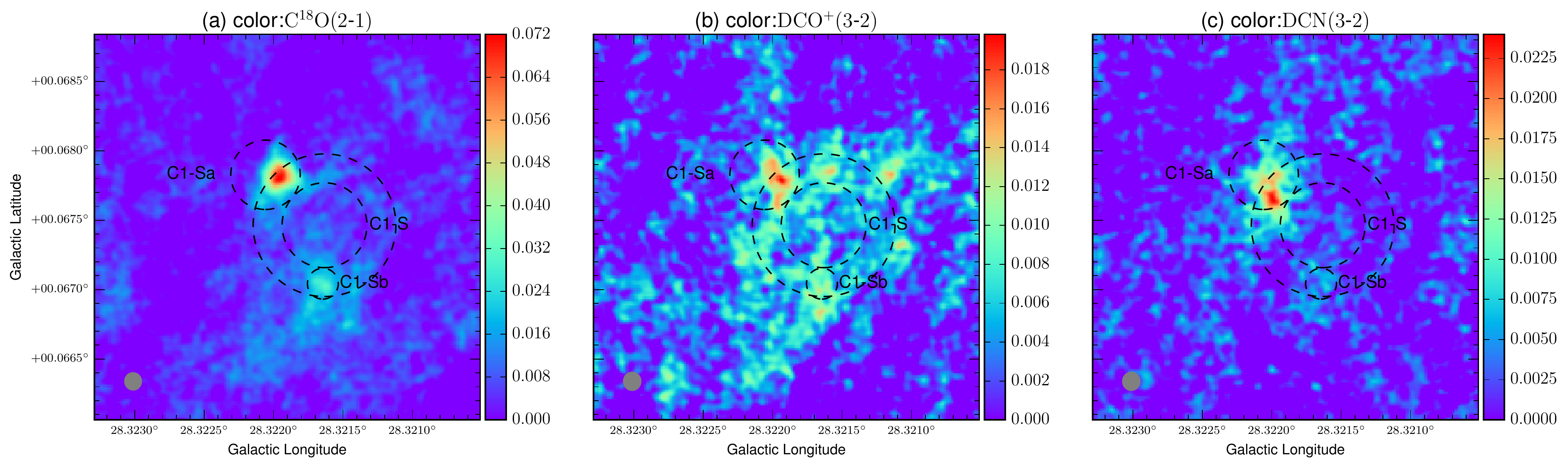}
\caption{
From left to right: 0th-moment maps of \ceions(2-1), \dcopns(3-2) and
DCN(3-2). All panels show the combined data.  
The integrations are from 76.9 to 81.9 \kmsns. The color
scale bars have unit of \jyb \kmsns. The synthesized beams ($\sim$
0.3\arcsec) are shown as gray filled ellipses in the lower-left
corners.
\label{fig:othermom0}}
\end{figure*}

\begin{figure*}[htb!]
\epsscale{1.}
\plotone{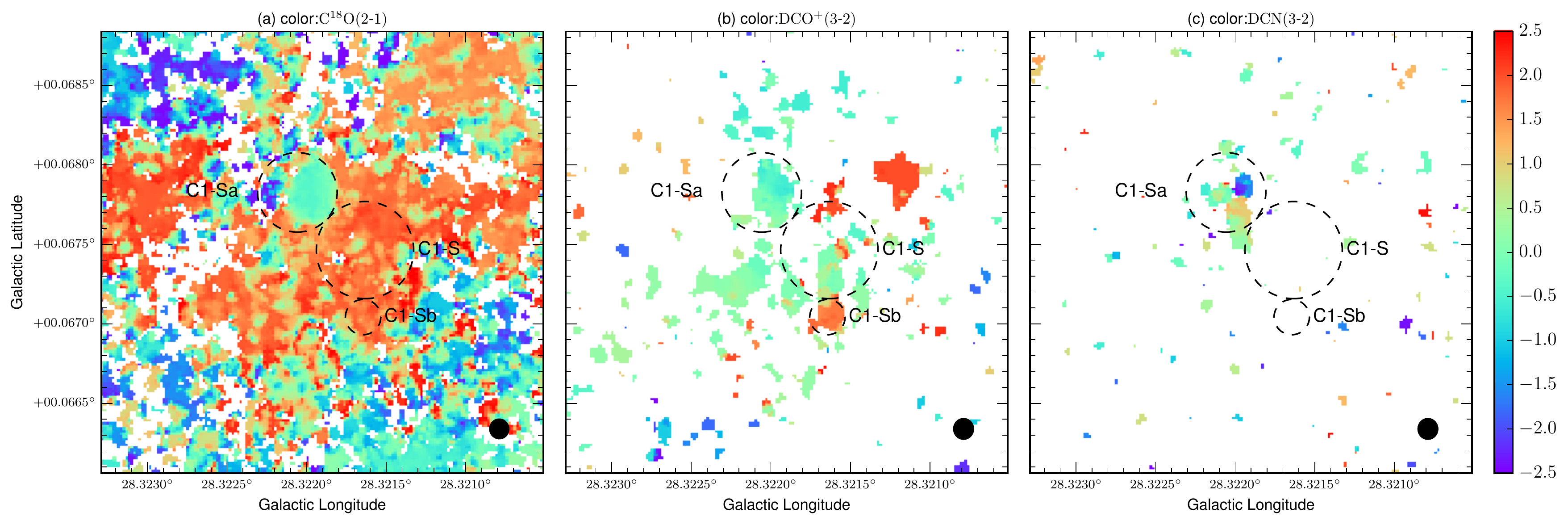}
\caption{
From left to right: 1st-moment maps of \ceions(2-1), \dcopns(3-2) and
DCN(3-2).  The integration is over 79.4$\pm$2.5 \kmsns.  Only cells
above 3$\sigma$ are considered.  The 1st-moment is then normalized to
the rest frame of 79.4 \kmsns.  The beams are shown as the black filled
ellipses in the lower-right corners.  The colorbar has the unit \kmsns.
\label{fig:othermom1}}
\end{figure*}

Figure \ref{fig:othermom0} shows the 0th-moment maps of \ceions(2-1),
\dcopns(3-2), and DCN(3-2).  The integrations are centered on 79.4
\kms and within a range from 76.9 to 81.9 \kmsns.  Note in Figure
\ref{fig:spec1} there is another velocity component at about 75 \kmsns
seen in \ceions(2-1), but this is not included in these maps.
\ceions(2-1) emission is nearly absent in C1-S (see also the spectra
shown in Figure \ref{fig:spec1}), but is rather mostly concentrated in
C1-Sa and C1-Sb.

The lack of \ceions(2-1) emission in C1-S can be explained by CO
depletion from the gas phase. In the C1-S core and much of its clump
envelope, we expect that the temperature is $\lesssim15\:$K (see,
e.g., Figure~\ref{fig:tkin}).
At such temperatures, CO is mostly frozen-out onto dust grains
\citep{2012A&ARv..20...56C}.  As shown in Figure~\ref{fig:spec1}(3c),
C1-S inner has no detected \ceions(2-1) flux (i.e., at velocities
immediately around 79.4~\kms overlapping with the $\rm N_2D^+$(3-2)
emission). We use the 3$\sigma$ 0th-moment \ceions(2-1) noise
level (0.446 K \kms for C1-S inner) to calculate an upper limit of
\ceio column density $N_{\rm C^{18}O}^{\rm
  C1-S~inner}<3.00\times10^{14}\:{\rm cm}^{-2}$, assuming an
excitation temperature of 10~K and optically thin conditions. This
corresponds to an abundance constraint of
[\ceions]/[H]$=2.91\times10^{-10}$ (using $\Sigma_{\rm c,mm,tot}$ from
Table \ref{tab:2} and assuming 1.0 g cm$^{-2}$ corresponds to $N_{\rm
  H}$ = 4.27$\times$10$^{23}$ cm$^{-2}$. Assuming [$^{12}$CO/H] =
10$^{-4}$ and [$^{16}$O/$^{18}$O] = 557 \citep{1999RPPh...62..143W},
the above abundance constraint in C1-S inner corresponds to a CO
depletion factor of $f_{D}> 616$. Such high CO depletion factors have
also been found toward the center of the lower-mass pre-stellar core
L1544 \citep{1999ApJ...523L.165C,Casellietal2002}, suggesting
similarity of astrochemical properties of this source with C1-S.

\ceions(2-1) emission is observed towards the protostars C1-Sa and
C1-Sb. This is to be expected, since in the protostellar models of
\citet{2014ApJ...788..166Z,2015ApJ...802L..15Z}, early-stage
protostellar cores can have mean temperatures up to $T\sim30$~K, which
is significantly above the CO sublimation temperature of 17~K
\citep{1993prpl.conf..163V}.

The 0th-moment map of DCO$^+$(3-2) shows emission from C1-Sa, a
relative lack of emission from C1-S inner (consistent with the large
amount of CO freeze-out), and hints of emission from the envelope
around C1-S inner. DCN(3-2) shows strong emission only from the C1-Sa
protostar, though with enhancement slightly offset from the peak
  of the continuum emission.

Figure \ref{fig:othermom1} shows the 1st-moment maps of \ceions(2-1),
\dcopns(3-2) and DCN(3-2), evaluated using the same velocity range as
the 0th-moment maps shown in Figure~\ref{fig:othermom0}. Only cells
above 3$\sigma$ values are included.  The \ceions(2-1) 1st-moment map
clearly shows two velocity components.  Much of the region is filled
with the 81.2 \kms component, while in C1-Sa the 79.4 \kms component
dominates.  

A plausible overall picture of the C1 region is the following.  There
are two overlapping clouds, one at 79.4 \kms and the other at 81.2
\kmsns.  The first one has strong CO depletion (except in a local
  region around the protostar C1-Sa)  and deuteration (especially with
\ntdpns), while the latter still has abundant CO but little \ntdpns.
C1-Sa and C1-S are in the 79.4 \kms
cloud, while C1-Sb is in the 81.2 \kms cloud.  While C1-S shows a
narrow line width in \ntdpns(3-2), C1-Sa line profiles are broader
(i.e., dispersion of 0.32~km/s from an HFS fit, 1.3 times larger than
that of C1-S inner), perhaps being caused by its outflows (Figure
\ref{fig:spec1}).  It is unclear if the 79.4 \kms and 81.2 \kmsns
clouds are interacting or not.

\subsection{Astrochemical Modeling of C1-S}

We carry out an astrochemical model of C1-S inner given its observed
properties. The chemical model is from \citet{2015ApJ...804...98K}: it
follows the time-dependent change of more than 100 species, including
\ntdpns. We adopt $n_{\rm H}=6\times 10^6\:{\rm cm}^{-3}$, i.e., the
envelope-subtracted estimate based on mm continuum emission. We set
gas and dust temperatures to 10~K and the heavy element depletion
factor to 600 (based on \ceio abundance estimation in
\S\ref{subsec:anci}).  Note we are assuming this depletion factor for
the elemental abundances of Carbon, Oxygen, and Nitrogen. The model
results show that the equilibrium abundance of \ntdp is [\ntdpns]/[H]
= 1.94$\times$10$^{-11}$.  If we decrease/increase $n_{\rm H}$ by a
factor of 2, the equilibrium value of [\ntdpns]/[H] = (1.84,
1.90)$\times$10$^{-11}$, respectively.  
If we set $T$ = (7, 13) K, the equilibrium [\ntdpns]/[H] =
(1.70, 2.10)$\times$10$^{-11}$, respectively.  
If we decrease/increase $f_{D}$ by a factor of 2, the
equilibrium [\ntdpns]/[H] = (3.11, 0.94)$\times$10$^{-11}$, respectively.  
If we decrease/increase cosmic ionization rate by a factor of 2, the
equilibrium [\ntdpns]/[H] = (1.91, 1.84)$\times$10$^{-11}$, respectively.  
These results illustrate the sensitivity of the theoretical
equilibrium abundance of \ntdpns to astrochemical model parameters.

On the observational side, if we assume an \ntdpns(3-2) excitation
temperature \tex = 10 K, the C1-S inner \ntdp column density is
1.39$\times$10$^{12}$ cm$^{-2}$.  Using the $\Sigma_{\rm
  c,mm}=1.61\:{\rm g\:cm^{-2}}$
from Table \ref{tab:2}, we derive an \ntdp abundance [\ntdpns]/[H] =
2.02$\times$10$^{-12}$. Considering the possibility of sub-thermal
excitation, we also try \tex = 6.4 K \citep{Fontani2011}, which yields
[\ntdpns]/[H] = 4.90$\times$10$^{-12}$.  For \tex = 4.1~K, which was
the low end of the range considered by \citet{2016ApJ...821...94K},
[\ntdpns]/[H] = 2.82$\times$10$^{-11}$.  

Given these results, we conclude that if \tex $\simeq 4\:$K, then the
observed abundance of \ntdp in C1-S inner is consistent with the
equilibrium value predicted by our astrochemical model for the
observed density, temperature and depletion factor. The timescale to
reach 90\% of the equilibrium is 1.36$\times$10$^{5}$~yr for an initial
ortho-to-para ratio of $\rm H_2$ of 3.
This is about a factor of 8 longer than the free-fall time of C1-S
inner. If the initial ortho-to-para ratio of $\rm H_2$ is 0.1,
the timescale will be shortened by a factor of 1.3 \citep{2015ApJ...804...98K}.

\section{Discussion and Conclusions}\label{S:discussion}

\begin{figure*}[htb!]
\epsscale{1.2}
\plotone{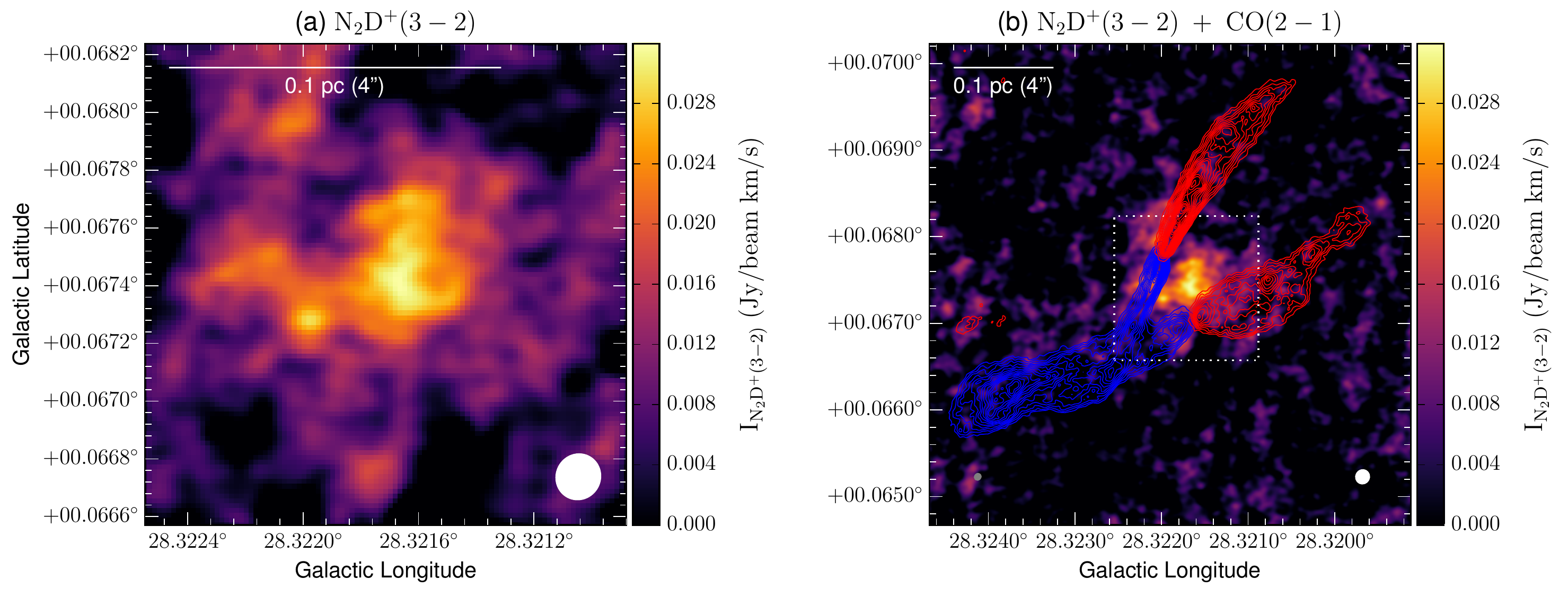}
\caption{
(a) Zoom-in view of Figure \ref{fig:ntdp1}c.
The synthesized beam ($\sim$ 0.5\arcsec) is shown as a white
filled ellipse at lower-right. 
(b) CO(2-1) 0th-moment contours overlaid on Figure \ref{fig:ntdp1}c.
The contour map is the same as Figure \ref{fig:co}b, except for 
starting from 15$\sigma$. 
The synthesized beam is shown in the lower right corner as a filled
white ellipse.
The dotted square shows the zoom-in region of panel (a).
\label{fig:summary}}
\end{figure*}

We have presented higher angular resolution follow-up observations of
the C1-S core, first identified by its \ntdpns(3-2) emission by T13. The
CO observations of T16 identified two protostellar outflow sources
near C1-S and it was speculated that one of these, C1-Sa, may be
forming from the core. However, now with about 10$\times$ better
angular resolution we are able to see that C1-S, as traced by its
\ntdpns(3-2) emission, is distinct from the C1-Sa and C1-Sb
protostars. Thus C1-S remains a good candidate to be a massive
($\sim50\:M_\odot$) starless core, being highly deuterated and CO
depleted ($f_D\gtrsim600$) and lacking concentrated mm continuum
emission and any sign of CO outflows 
 (see the summary Figure \ref{fig:summary}).

The velocity dispersion of C1-S, as traced by \ntdpns(3-2), indicates
either sub-virial conditions in the trans-Alfv\'enic case, or else
that the core is in approximate virial equilibrium with stronger
magnetic fields ($\sim2\:$mG) that dominate over turbulent pressure
support, i.e., turbulence is sub-Alfv\'enic within the core. We favor
the latter interpretation, since such $B$-field strengths are
consistent with an extrapolation of the $B$ versus density relation of
\citet{2010ApJ...725..466C}. They also help explain the mass of C1-S
at its outer scale as being set by the magnetic critical
mass. Astrochemical modeling indicates that a relatively long core age
compared to the free-fall time is needed to reach the observed highly
deuterated state, and once a core is older than a few free-fall times,
it is expected to be able to approach approximate virial
equilibrium. Predictions of this scenario are that the continuum
emission from C1-S will show strong polarization and that Zeeman
splitting may be observed from species such as CN, if they remain in
the gas phase at a level high enough to be observed. Unfortunately the
weakness of the dust continuum emission and the apparent high heavy
element depletion factor in C1-S may make these tests difficult to
carry out. Another prediction is that infall velocities, i.e., as
traced by inverse P-Cygni profiles from suitable species \citep[e.g., the
\nthpns(1-0) line, as has been observed in
lower-mass pre-stellar cores;][]{2010MNRAS.402.1625K}, will show infall
speeds that are a small fraction of the local free-fall speed.

These observations break new ground in being able to resolve the
kinematic structure of the massive starless core. We have been able to
measure the radial profile of turbulence in the core, which is seen to
have a relatively constant velocity dispersion with radius. We have
presented high resolution first moment maps of the internal velocity
structure. The statistics of these kinematic properties can be
compared against numerical simulations of such cores
\citep{2016ApJ...833..274G}. We defer a detailed comparison with such
simulations to a future paper.

The observations presented here have also revealed more details about
the C1-Sa and C1-Sb protostars. C1-Sa is consistent with being a
massive protostellar core ($\sim30\:M_\odot$) that is in a very early
stage of formation. It exhibits a very collimated outflow with a
half-opening angle of $\lesssim10^\circ$.
The presence of \ntdp in C1-Sa is consistent with the findings by
\citet{2009A&A...493...89E}, who found that the \ddfrac
can be as large as 20\% in young low-mass protostars, 
while it declines in more evolved sources. 
C1-Sb appears to be a lower mass protostellar core ($\sim2\:M_\odot$),
but at a later stage of development, e.g., with a wider outflow cavity
opening angle.

The overall presence of a massive starless core alongside a relatively
massive protostellar core and lower mass protostellar core gives
interesting clues about the early stages of massive star formation, at
least in one case. The environment within this region of about
  0.1~pc projected scale, is not especially crowded, particularly
considering that C1-Sb seems to be at a quite different velocity and
may be physically separated along the line of sight. This leaves just
the C1-S and C1-Sa sources, both relatively massive, monolithic
objects. There is no evidence so far for a cluster of protostars that
all drive outflows, which is an expectation of Competitive Accretion
models \citep[e.g.,][]{2001MNRAS.323..785B,2010ApJ...709...27W}. 
The current observations can easily detect sources like C1-Sb that
have about a $2\:M_\odot$ protostellar core and its level of
protostellar outflow activity, if they are present. The $5\sigma$
continuum mass sensitivity is $0.15\:M_\odot$ per beam (for
$T=20\:$K).

Deeper, more sensitive observations with {\it ALMA} will lead to
improved understanding of these sources. For example, higher
sensitivity \ntdp observations can better probe the kinematic
subtructure of the C1-S core. Other transitions should be observed to
constrain the excitation temperature of this species. Similar
observations of \nthpns are needed to map the deuteration
structure. Higher sensitivity and angular resolution observations in
high velocity CO(2-1) and in mm dust continuum will enable more
stringent constraints to be placed on the protostellar content of this
region. A search for cm continuum emission is needed to better
constrain models \citep{2017ApJ...835...32T} of the potentially massive
protostar C1-Sa.

\acknowledgments We thank an anonymous referee for comments that 
improve the paper. We thank Mengyao Liu, Matthew Goodson, Crystal Brogan, 
Jim Braatz, Sarah Wood, Shawn Booth, and H\'ector Arce for helpful discussions. 
SK was funded by NSF award AST-1140063 while conducting this study.
JCT acknowledge an NRAO/SOS grant and NSF grant
AST1411527.  PC acknowledges the financial support of the European
Research Council (ERC; project PALs 320620).  This paper makes use of
the following ALMA data: ADS/JAO.ALMA\#2013.1.00248.S. ALMA is a
partnership of ESO (representing its member states), NSF (USA) and
NINS (Japan), together with NRC (Canada), NSC and ASIAA (Taiwan), and
KASI (Republic of Korea), in cooperation with the Republic of Chile.
The Joint ALMA Observatory is operated by ESO, AUI/NRAO and NAOJ.  The
National Radio Astronomy Observatory is a facility of the National
Science Foundation operated under cooperative agreement by Associated
Universities, Inc.

{\it Facilities:} \facility{ALMA}; \facility{VLA}

\bibliography{ref}
\bibliographystyle{aasjournal}

\end{document}